\documentclass[twocolumn,twoside]{IEEEtran}

\ifCLASSINFOpdf
  
\else
 
\fi

\ifCLASSOPTIONcompsoc
\usepackage[caption=false, font=normalsize, labelfont=sf, textfont=sf]{subfig}
\else
\usepackage[caption=false, font=normalsize]{subfig}
\fi
\usepackage{lipsum}%
\usepackage[dvipsnames]{xcolor}

\usepackage{balance}
\usepackage{multicol}   
\usepackage{cite}
\usepackage{gensymb}
\usepackage{multirow}
\usepackage{graphics}
\usepackage{epsfig}
\usepackage{graphicx}
\usepackage{epstopdf}
\usepackage{textcomp}
\usepackage{amsmath}
\usepackage{mathtools}
\usepackage{filecontents}
\usepackage{lipsum,color}
\usepackage{amssymb}
\usepackage{float}
\usepackage{times} 
\usepackage{amsthm}  

\usepackage{amsfonts}

\theoremstyle{break}

\begin{document}
\title{Modulating Retroreflector-based Satellite-to-Ground Optical Communications: Sensing and Positioning}

\author{Mohammad~Taghi~Dabiri,~Mazen~Hasna,~{\it Senior Member,~IEEE},\\
	~Saud~Althunibat,~{\it Senior Member,~IEEE},~and~Khalid~Qaraqe,~{\it Senior Member,~IEEE}
\thanks{Mohammad T. Dabiri and Mazen Hasna are with the Department of Electrical Engineering, Qatar University, Doha, Qatar (e-mail: m.dabiri@qu.edu.qa; hasna@qu.edu.qa).}
\thanks{S. Althunibat is with the Department of Communication Engineering, Al-Hussein Bin Talal University, Ma’an 23874, Jordan (e-mail:
	saud.althunibat@ahu.edu.jo).}
\thanks{K. Qaraqe is with the Department of Electrical and Computer Engineering, Texas A\&M University at Qatar, Doha 23874, Qatar (e-mail:
	khalid.qaraqe@qatar.tamu.edu).}	
\thanks{This publication was made possible by grant number NPRP14C-0909-210008 from the Qatar National Research Fund, QNRF. }
}

\maketitle
\begin{abstract}
This paper focuses on the optimal design of a modulated retroreflector (MRR) laser link to establish a high-speed downlink for cube satellites (CubeSats), taking into account the weight and power limitations commonly encountered by these tiny satellites. To this end, first, a comprehensive channel modeling is conducted considering key real channel parameters including mechanical gimbal error, fast steering mirror angle error, laser beamwidth, MRR area, atmospheric turbulence, and channel coherence time. Accordingly, a closed-form expression for the distribution of the received signal is derived and utilized to propose a maximum likelihood based method to sense and estimate the initial position of the satellite. Subsequently, the distribution of the distance estimation error during the sensing phase is formulated as a function of the laser beamwidth and the gimbal error, which enables us to fine-tune the optimal laser beamwidth to minimize sensing time. Moreover, using the sensing and initial satellite distance estimation, two positioning algorithms are proposed. To compare the performance of the proposed positioning method, we obtain the lower bound of the positioning error as a benchmark. Finally, by providing comprehensive simulations, we evaluate the effect of different parameters on the performance of the considered MRR-based system in both the sensing and positioning phases.
\end{abstract}

\IEEEpeerreviewmaketitle

\section{Introduction}
\IEEEPARstart{C}{ube} Satellites (CubeSats) are tiny box-shaped miniaturized satellites with a form factor of 10 cm cubes and a mass of no more than 2 kg. This type of small satellites has rapidly become a popular platform which is mainly launched into low Earth orbit (LEO) \cite{saeed2020cubesat}.
The global CubeSat market size was valued at 210 million in 2021 and it is expected to reach USD 857 million by 2030 \cite{2030_cube_market}.
However, due to the constraints on the size dimensions which leads to a small surface area on their external walls, CubeSats have a small harvestable solar power. Moreover, launch providers may impose limits on the battery capacity of CubeSats due to safety regulations \cite{NASA_nanoRack}. Therefore, the available power for the communication payload is limited to about 2 W \cite{ochoa2014deployable}. As there is a direct relationship between transmitted power and receiver signal-to-noise ratio (SNR), conventional RF links considered for the downlink of CubeSats have a limited capacity. 
In addition, due to the spectrum limitations as well as the limitations related to the dimensions of RF antennas, the downlink rate of CubeSats is limited to a range from tens of kbps to a few Mbps. For example, one of the high-speed transmitters available in the market for CubeSats can achieve a maximum data rate of 4.3 Mbps in the S-band \cite{4Mbps_Tx_marjet}. Such rates are notably lower than the several hundreds of Mbps to several Gbps typically achieved by traditional satellites.

Due to unique features like high unlicensed bandwidth, anti-interception, compact size, and low power consumption, free-space optical (FSO) communication, utilizing laser technology, has gained significant appeal in space communications and serves as a cornerstone for the development of space networks \cite{li2021advanced,kaushal2016optical,khalighi2014survey}. Hence, there has been a significant surge in recent efforts towards the advancement of laser communications among satellites \cite{toyoshima2021recent,carrasco2020space,chaudhry2022temporary,chaudhry2020free,le2021throughput,le2023fso,walsh2022demonstration}.
The focus of all these works is on the design of common links in such a way that the laser payload includes a receiver (for uplink) and a transmitter (for downlink).














The main challenge of CubeSat communication lies in the downlink due to two main reasons. First, it needs power supply for data transmission \cite{carrasco2022development}, and second, the laser link requires additional hardware to provide precise alignment at the transmitter side \cite{ishola2022characterization,kolev2023latest}.
%
For a satellite at an altitude of 500 km, an alignment error of even 5 \textit{mrad} causes the laser beam center to deviate by 2,500 m. To compensate for this deviation, a larger laser beamwidth should be selected which causes a small portion of the transmitted power to be collected by the receiver aperture, and thus, more transmitted power is needed to compensate for this loss. Therefore, together with the transmitter system, compact stabilizer and aligner systems must be designed and installed, which, while increasing the cost and weight, occupy almost the entire space of the cube and a little space is left for other equipment such as cameras with stronger lenses.

\begin{figure*}
	\centering
	\subfloat[] {\includegraphics[height=3.2 in]{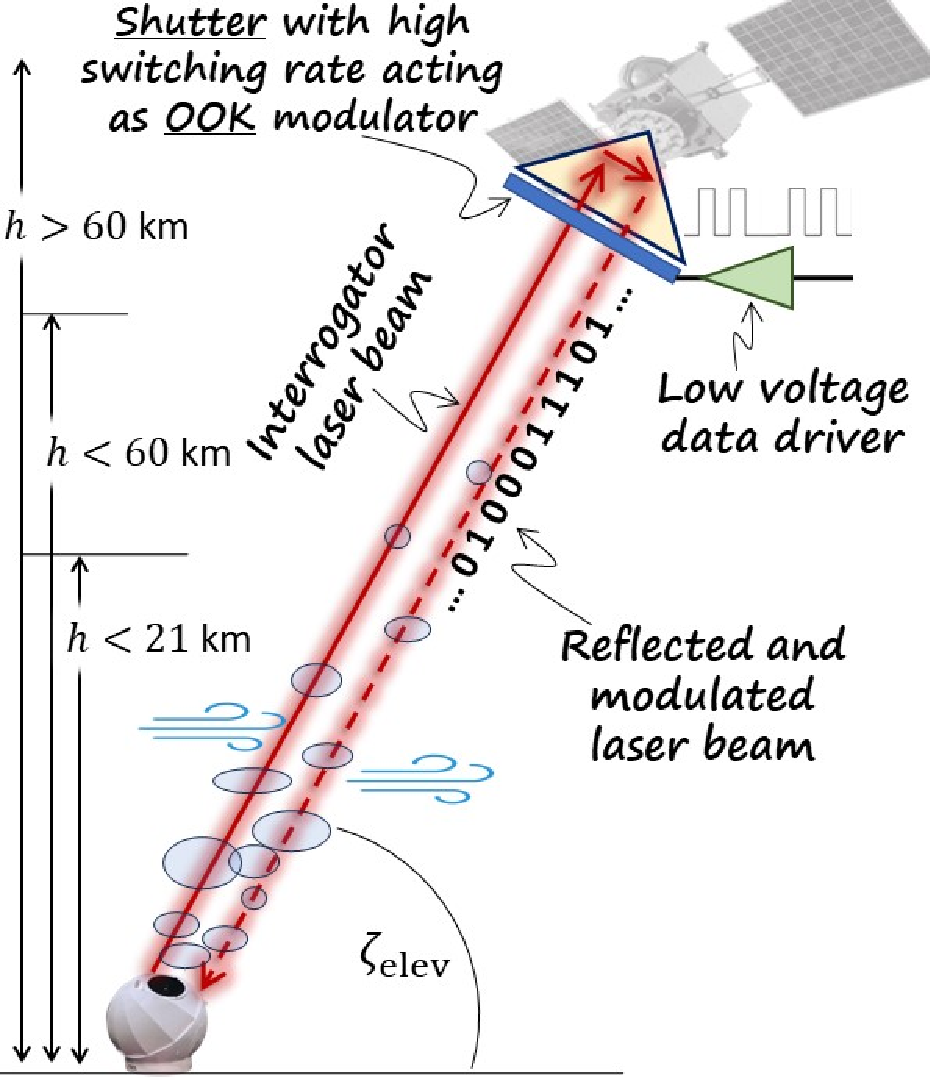}
		\label{sesf1}
	}
	\hfill
	\subfloat[] {\includegraphics[height=3.2 in]{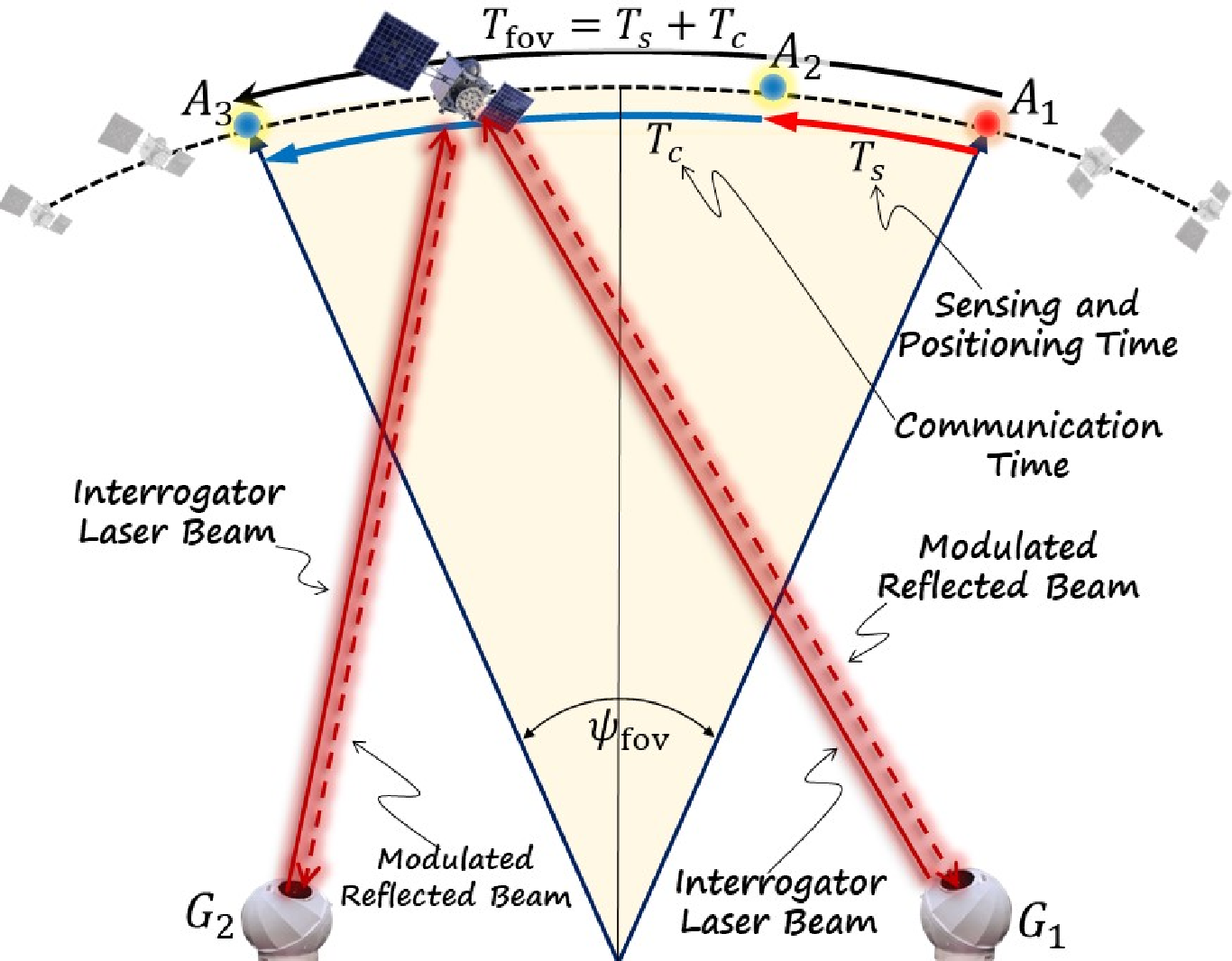}
		\label{sesf2}
	}
	\caption{(a) Visual representation depicting the establishment of an MRR-based laser link, wherein the interrogator laser signal is modulated and reflected to the GS without requiring power and stabilizers in the satellite. (b) Given the satellite's rapid movement, limited time of availability, and potential tracking errors in the GS gimbal system, there is a crucial need for swift and precise sensing and positioning to reduce the geometrical pointing loss and create optimal conditions for establishing a high-data-rate downlink.
	}
	\label{sesf}
\end{figure*}

As an alternative solution, a retroreflector-based topology can be employed for downlink \cite{trinh2021experimental, born2018all, quintana2021high}. A retroreflector is a passive element suitable for asymmetric scenarios where one of the nodes is constrained by weight and power limitations. For downlink scenario, the interrogator laser signal is pointed from the ground station (GS) toward the modulating retroreflector (MRR) array installed on the satellite as shown in Fig. \ref{sesf}. Due to the perpendicular feature of the mirrors to each other,
the collected interrogator laser signal by MRR is modulated (by using a shutter in front of the MRR aperture) and directly reflected to the GS. Such a topology is usually considered passive as it does not transmit power. Moreover, in this topology, there is no need to equip the satellite with aligning devices, which significantly reduces the weight, complexity, and cost.
However, the main challenge of this design is that the shutter used in front of the MRR aperture has a low modulation rate, thereby, most of the retroreflector-based applications are limited to laser ranging \cite{degnan2023tutorial,moon2023performance}.
%


Given the rapid technological developments, shutter modulation rates have recently reached close to Gbps \cite{trinh2021experimental, born2018all, quintana2021high} and they are expected to reach higher modulation rates in the future. To this end, the analysis and designing of Gbps communication using MRR array has been the subject of \cite{trinh2021experimental,quintana2021high}. However, these analysis and implementations have been tested in drone dimensions.
To design the considered high data rate MRR-based downlink for CubeSat, it is necessary to pay attention to three important challenges. First, unlike the conventional links, here, we have a long round-trip path that doubles the atmospheric attenuation and also increases the intensity of the atmospheric turbulence \cite{moon2023pointing}.
Second, unlike the laser ranging systems which have a frequency in the order of KHz, for the considered MRR-based downlink, the received SNR decreases by increasing the data rate. For instance, for a Gbps downlink, the pulse width reduces to ns, which decreases the SNR by $\simeq 10^6$ times compared to a KHz laser ranging. Third, the shutter modulation rate is inversely related to the active area of the shutter which is denoted by $A_\text{MRR}$, and to reach higher data rates, $A_\text{MRR}$ must be reduced ($A_\text{MRR}<0.0001$ m$^2$). In this case, the small $A_\text{MRR}$ along with the alignment errors can cause more than 200 dB loss. This loss is often called geometrical pointing loss. 
For ideal conditions (assuming zero alignment error), as the laser beamwidth decreases, the geometrical pointing loss decreases.
However, in real conditions, a small beamwidth makes the system more sensitive to the alignment error and causes more attenuation. To design a high data-rate MRR-based laser system, a very accurate alignment system is required, which itself is a function of the mechanical errors of the gimbal system, the errors of fast steering mirrors (FSMs), and atmospheric turbulence-induced beam wandering. LEO satellites are orbiting the Earth at a speed of about 7-10 km/s and within a short period, they are moving in the field of view (FoV) of the GS. Therefore, in addition to high alignment accuracy, as shown in Fig. \ref{sesf2}, the sensing and positioning time of the satellite is also of great importance. A key point for designing the considered system is that all the mentioned factors such as mechanical gimbal errors, errors of FSMs, laser beamwidth, sensing time, positioning accuracy, and data rate are in a chain and highly interdependent. Nevertheless, none of the existing literature comprehensively analyzes and models all these parameters together. In addition, the joint positioning, tracking, and communication methods provided in \cite{bashir2021optimal,tsai2023angle,bashir2020adaptive} are not applicable for the considered MRR-based topology.

To this end, the optimal design of an MRR-based laser link to create a high-speed downlink for CubeSat is the main subject of this work. In particular, considering the importance of alignment accuracy, we focus on the analysis and design of the sensing and positioning systems and study the relationship between the system parameters such as mechanical gimbal error and FSM alignment error with sensing time and positioning accuracy. Finally, we obtain the optimal laser beamwidth to reduce the sensing time and increase the positioning accuracy. The contributions of this paper are summarized as follows:
\begin{itemize}
	\item First, a comprehensive modeling of the round trip channel between the CubeSat and the GS is performed by taking into account the real channel parameters such as mechanical gimbal error, FSM angle error, laser beamwidth, MRR area, atmospheric turbulence, atmospheric loss, and channel coherence time. 
	\item Second, we derive a closed-form expression for the distribution of the received signal as a function of the channel parameters. We demonstrate that gimbal alignment errors on the order of \textit{mrad} result in a significant geometrical pointing loss, making direct satellite positioning unfeasible.
	\item Third, an initial sensing phase is designed using Gaussian acquisition to sense the location of the satellite and make an initial estimate of the satellite location in the order of several tens of $\micro$rad. For initial estimation, we propose a maximum likelihood (ML)-based method. Later, to reduce the complexity of the ML-based method, we provide a time-averaging algorithm, which can be used as an input to reduce the complexity of the ML-based method. 
	\item Forth, we obtain the distribution of the distance estimation error in the sensing phase as a function of laser beamwidth, gimbal error, and the random location of the satellite in the ambiguity area. This distribution helps us to analytically obtain the average sensing time as a function of tunable parameters such as the laser beamwidth. The distribution expression is employed to dynamically adjust the optimal value of the laser beam based on the prevailing channel conditions, aiming to minimize the sensing time. 
	\item In the subsequent step, leveraging the sensing data and the initial satellite distance estimation, we proceed to the positioning phase, and two positioning algorithms are proposed. Both of them are a function of the initial estimation error in the sensing phase as well as the FSM errors. 
	We demonstrate that the optimal laser beamwidth during the positioning phase is contingent upon the initially estimated distance obtained in the sensing phase.
	In addition, to compare the performance of the proposed positioning method, we obtain the lower bound of the positioning error as a benchmark.
	\item Finally, by providing comprehensive simulations, we evaluate the effect of different parameters on the performance of the considered MRR-based system in both phases of sensing and positioning.
\end{itemize}

\section{The System Model}
In this work, we consider a communication system based on a set of CubeSats in such a way that each satellite collects information and images during its orbital movement and sends them to one or more GSs. As mentioned earlier, the most important challenge of CubSats is the small dimensions to use high gain RF antennas, lower transmission power, and spectrum limitations that limit the transmission rate in the order of several hundred kbps to several Mbps. To overcome this limitation and create satellite downlink with high data rate, we use laser and MRR technologies. The MRR array is installed on the satellite's body. As shown in Fig. \ref{sesf}, when the satellite enters the FoV of the GS, the interrogator laser signal is pointed towards the satellite. Due to the fact that the MRR mirrors are perpendicular to each other, the collected laser signal is directly reflected to the GS. To modulate the reflected signal, a shutter with high switching rate is used in front of the cube retroreflector which only requires a low voltage data driver. To send bit "0", the shutter acts as an insulator and does not reflect any signal, while for bit "1", it reflects a high percentage of the received signal. In other words, it modulates the optical intensity similar to on-off keying (OOK) modulation.

\subsection{Channel Modeling}
The considered link consists of multiplying two channel coefficients of the round-trip optical channel. In addition, the data rate of MRR is inversely proportional to the MRR's aperture area $A_\text{MRR}$, and to achieve higher rates, it is necessary to choose a small $A_\text{MRR}$ (on the order of $A_\text{MRR}<0.0001$ m$^2$). Therefore, it is necessary to use an MRR array to increase the SNR similar to \cite{10299800}.
Therefore, at any discrete time $k$, the received signal at the GS is modeled as follows:
\begin{align}
	\label{d1}
	P_r[k]  = R h[k] P_t s[k] + n[k],
\end{align}
where $P_t$ is the transmitted interrogator laser power, $s[k]$ is the transmitted OOK signal, $n[k]$ is the additive thermal noise with variance $N_0$, $R$ denotes the photo-detector responsivity, and channel coefficient $h$ is modeled as 
\begin{align}
	\label{ch1}
     &h[k] = h_{L1}h_{L2}  h_{pg}   \\&~~~\times \sum_{m=1}^M h_{ps}[k,m]   h_\text{MRR}[k,m,\theta_\text{MRR}]
     h_{a1}[k,m]h_{a2}[m,k], \nonumber
\end{align}
where $M$ is the number of the MRRs, and $h_\text{MRR}[k,m,\theta_\text{MRR}]$ represents the fraction of power directly reflected to the total power collected by the $m$th MRR, which is a function of the angles between the incident laser beam and the vector of the $m$th MRR denoted by $\theta_\text{MRR}$ (for more details please see \cite[Fig. 5]{dabiri2022modulating}).
For $\theta_\text{MRR}$ lower than 2 degrees, more than 90\% of the signal is reflected \cite{dabiri2022modulating}. Since for satellite communication, alignment accuracies in the order of less than \textit{mrad} is needed, we can well assume $h_\text{MRR}\simeq 1$. By doing this, \eqref{ch1}, is simplified as:
\begin{align}
	\label{ch2}
	&h[k] = h_{L1}h_{L2}  h_{pg}   \sum_{m=1}^M h_{ps}[k,m]  
	h_{a1}[k,m]h_{a2}[m,k],
\end{align}
where the parameters in \eqref{ch2} are modeled in the following.
$h_{L1}$ and $h_{L2}$ are the atmospheric attenuation of GS-to-satellite and satellite-to-GS links, respectively, and are typically modeled by the Beer-Lambert law as \cite{ghassemlooy2019optical}:
\begin{align}
	\label{ch3}
	h_{L1}=h_{L2}=h_L=e^{-Z \zeta},
\end{align}
where $Z\simeq \frac{H_s}{\sin(\zeta_\text{elev} )}$ is the linklength, $\zeta_\text{elev}$ is the elevation angle, $H_s$ is the satellite height, and $\zeta$ is the scattering coefficient which is a function of visibility. $h_{L1}$ and $h_{L2}$ are equal since the parameters $\zeta$ and $Z$ are the same for both links.

The Gamma-Gamma (GG) distributions is a good candidate to efficiently model weak to strong ranges of atmospheric turbulence conditions \cite{ghassemlooy2019optical,andrews2005laser} and is formulated as
\begin{align}
	\label{ss8}
	f_G\left(h_{ai}\right) = \frac{2(\alpha\beta)^{\frac{\alpha+\beta}{2}}}{\Gamma(\alpha)\Gamma(\beta)}
	h_{ai}^{^{\frac{\alpha+\beta}{2}-1}}
	k_{\alpha-\beta}  \left( 2\sqrt{\alpha\beta h_{ai}}\right),
\end{align}
where $i\in\{1,2\}$, $\Gamma(\cdot)$ is the Gamma function and $k_m(\cdot)$ is the modified Bessel function of the second kind of order $m$. Also, $\alpha$ and $\beta$ are respectively the effective number of large-scale and small-scale eddies, which depend on Rytov variance $\sigma_R^2$ \cite{andrews2005laser}.
For two nodes with different heights, Rytov variance is obtained as \cite[p. 509]{andrews2005laser}:
\begin{align}
	\label{sss1}
	&\sigma_R^2 =   \frac{2.25 k_\ell^{7/6}}{\sin^{11/6}(\zeta_\text{elev})}  \int_{H_0}^{H_s} 
	C_n^2(Z_h) \left( Z_h-H_0 \right)^{5/6}   \textrm{d}Z_h,
\end{align}
where $H_0$ is the height of GS, and 
\begin{align}
	&C_n^2(Z_h) = 0.00594 (\mathcal{V}/27)^2 \left(10^{-5} Z_h \right)^{10}
	\exp\left( -\frac{Z_h}{1000} \right) \nonumber \\
	&~~~ + 2.7 \times 10^{-16}   \exp\left( -\frac{Z_h}{1500} \right) 
	+ C_n^2(0) \exp\left( -\frac{Z_h}{100} \right), \nonumber
\end{align}
is the refractive-index structure parameter at height $Z_h$ which characterizes the atmospheric turbulence, $\mathcal{V}$ (in m/s) is the speed of strong wind and $C_n^2(0)$ (in $\textrm{m}^{-2/3}$) is a strong nominal ground turbulence level.

In \eqref{ch2}, $h_{ps}[k,m]$ is the geometrical pointing loss of the $m$th elements of MRR array which is induced due to the tracking errors of the satellite.
Unlike conventional satellite optical links, for this topology, we have two round-trip paths that double the atmospheric attenuation and the intensity of atmospheric turbulence. Therefore, to compensate for these negative effects, we must reduce the geometrical pointing loss by using a higher alignment accuracy in the order of several $\micro$rad. In the following sections, with the modeling and technical analysis of geometrical pointing loss, we show that to achieve higher alignment accuracy, three different phases including (i) acquisition and sensing, (ii) positioning, and (iii) joint communication and tracking must be performed sequentially.
$h_{pg}$ is the geometrical loss of the GS receiver, and since the MRR reflects the laser signal at the same input angle, it can be well approximated as \cite{dabiri2022modulating}:
\begin{align}
	\label{dd2}
	h_{pg} \simeq {4 d_g^2}/{w_{zg}^2} =  {4 d_g^2}/{Z^2 \theta_\textrm{div}^2},
\end{align}
where $w_{zg}$ is the optical beamwidth in the GS, and $d_g$ is the aperture radius of the GS.

\section{Sensing and Positioning}

\subsection{Background}
According to the satellite's movement on a certain orbit, the instantaneous location of the satellite is known to the GS. When the satellite enters the FoV of the GS, the ground station starts tracking the satellite using a gimbal. The angular accuracies of the gimbal are in the order of several \textit{mrad} to degrees. In this work, we show that even high-precision gimbals close to one \textit{mrad} are not able to establish the considered MRR-based downlink, because the round-trip path of the laser signal, along with the high geometrical loss, reduces the received signal to below the noise level.

Let $\theta_{ge} \sim\mathcal{N}(0,\sigma_{\theta_{ge}}^2)$ represent the angle error of the gimbal and $R_{ge} \sim\mathcal{N}(0,\sigma_{ge}^2)$ represent the spatial error between the center of the laser beam and the satellite, where $\sigma_{ge}\simeq Z \sigma_{\theta_{ge}}$. 
If we only use the gimbal, a random distance $r_{ds}$ is created between the satellite and the laser beam center. Let $p_{g}=(0,0)$ denote the center of the laser beam caused by the gimbal and $p_s=(x_s,y_s)$ denote the location of the satellite in the $x-y$ coordinate plane perpendicular to the gimbal axis, where $x_s\sim\mathcal{N}(0,\sigma_{ge})$ and $y_s\sim\mathcal{N}(0,\sigma_{ge})$. The target is to accurately estimate $p_s$ with respect to $p_g$.
We consider a Gaussian beam at the GS, for which the normalized spatial distribution of the received intensity at distance $Z$ is given by \cite{saleh2019fundamentals}
\begin{align}
	I_r(d,Z) = \frac{2}{\pi w_z^2} \exp\left(-\frac{2(x^2 + y^2)}{w_z^2} \right),
\end{align}
where $d = [x, y]$ is the radial distance vector from the beam center. Also, $w_z$ is the beamwidth at distance $Z$ and can be approximated as $w_z\simeq \theta_\text{div} Z$, where $\theta_\textrm{div}$ is the optical beam divergence angle \cite{ghassemlooy2019optical}. Since the optimal value of $w_z$ changes in different phases, we use $w_{zs}$, $w_{zp}$, and $w_{zc}$, respectively, to denote the beamwidth in the sensing, positioning, and communication phases.
If only the gimbal is used for tracking, the geometrical pointing loss caused by each element of the MRR array (the ratio of the power collected by the $m$th MRR to the total power distributed in the laser beam) is obtained as:
\begin{align}
	\label{sb2}
	h_{ps}[k,m] &= \frac{2}{\pi w_{zs}^2} \int \int_{p_{A_m}(x,y)} \exp\left(-\frac{2(x^2 + y^2)}{w_{zs}^2} \right)  \textrm{d}x \textrm{d}y,
\end{align}
where $p_{A_m}(x,y)$ is the position of effective aperture area of the $m$th MRR in $x-y$ plane. Unlike terrestrial links, the laser beamwidth for satellite communications is on the order of several tens of meters. On the other hand, the dimensions of each MRR element are in the order of less than a centimeter, and the entire MRR array is in the order of a few centimeters. Therefore, with good accuracy, \eqref{sb2} is simplified as:
\begin{align}
	\label{sb3}
	&h_{ps}[k,m] =h_{ps}[k,m'] \simeq \frac{2A_\text{MRR}}{\pi w_{zs}^2}  \exp\left(-\frac{2(x_s^2 + y_s^2)}{w_{zs}^2} \right) ,
\end{align}
where $(m,m')\in\{1,...,M\}$ and $A_\text{MRR}$ is the MRR's aperture area.
Using \eqref{ch2} and \eqref{sb3}, the geometrical pointing loss due to the MRR array is obtained as:
\begin{align}
	\label{sb4}
	& h_{ps}[k] = \frac{1}{M} \sum_{m=1}^M h_{ps}[k,m] \simeq \frac{2A_\text{MRR}}{\pi w_{zs}^2}  \exp\left(-\frac{2(x_s^2 + y_s^2)}{w_{zs}^2} \right) .
\end{align}
Based on \eqref{sb4}, at $Z=500$ km and $\sigma_{\theta_{ge}}=5$ \textit{mrad}, we have $\sigma_{ge}=2500$ m, which can cause geometrical pointing coefficient even lower than -300 dB, and almost in most cases the reflected signal is below the noise level. To increase the alignment accuracy, simultaneously with gimbal tracking, fast steering mirrors (FSMs) are used, which have an alignment accuracy in the order of $\micro$rad. 
However, to achieve such accuracy using FSM, it must first have an accurate positioning of the satellite location. To this end, first, an angular space acquisition should be performed by FSMs around the axis of the gimbal.
During the acquisition process, as soon as the laser beam is near the satellite, the geometrical pointing loss is reduced and the power reflected from the satellite is sensed by the GS.
After satellite sensing, to achieve higher estimation accuracy, we enter the positioning phase. 

\begin{figure}
	\centering
	\subfloat[] {\includegraphics[width=1.65 in]{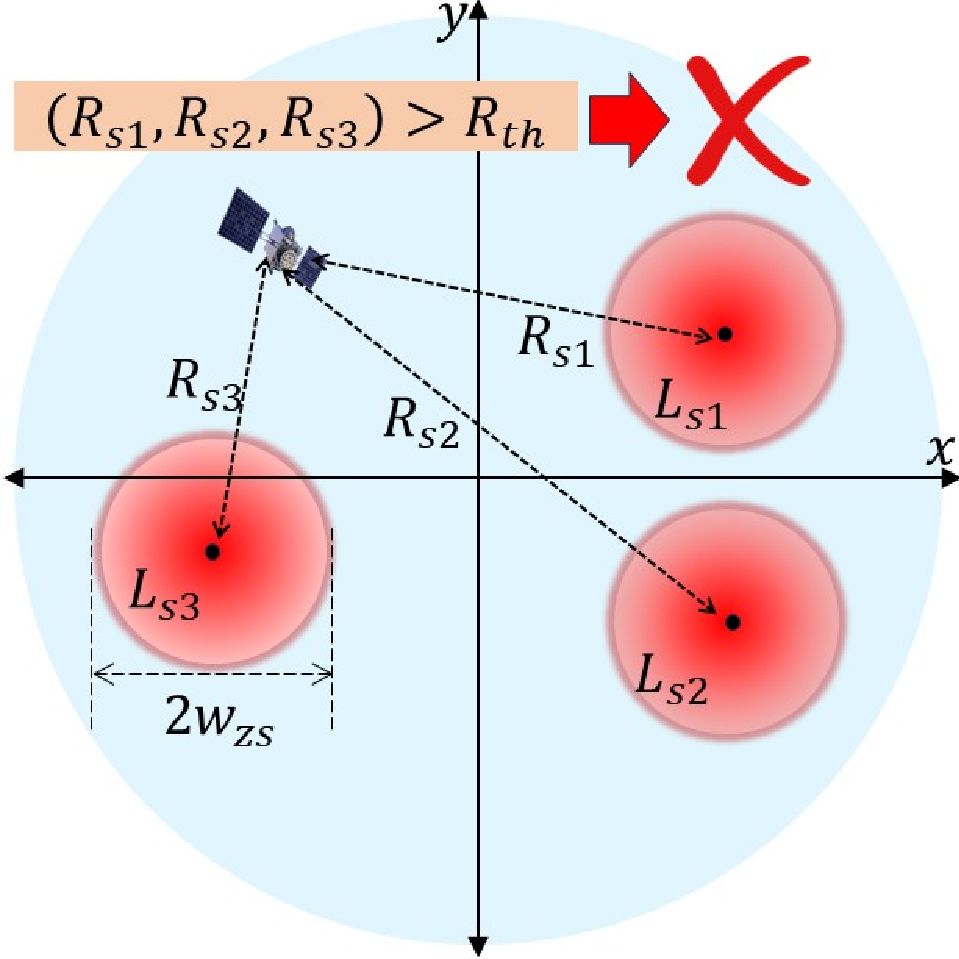}
		\label{sef1}
	}
	\hfill
	\subfloat[] {\includegraphics[width=1.65 in]{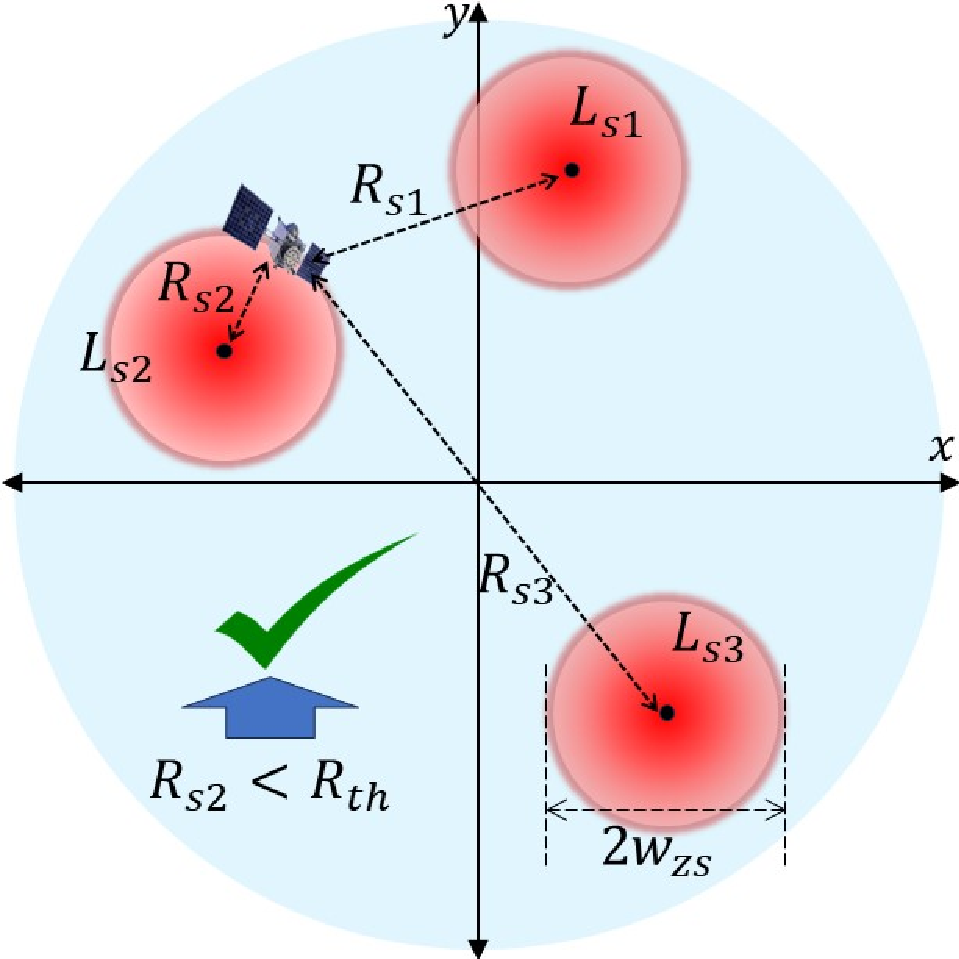}
		\label{sef2}
	}
	\caption{Two examples of the acquisition and sensing phase: (a) A random distribution of laser beams where none of the beams are located near the satellite and no signal is sensed by the GS; (b) The random distribution of beams is reiterated, until finally a beam is placed near the satellite, and thus, the received reflected power can be sensed by the GS.
	}
	\label{sef}
\end{figure}

\subsection{Sensing Phase}
In this work, we use $N_m$ FSM to perform the acquisition process around the gimbal axis in such a way that at each moment, each FSM creates a random angle $\theta_{Fi}=\theta'_{Fi}+\theta_{ei}=(\theta'_{Fxi}+\theta_{exi},\theta'_{Fyi}+\theta_{eyi})$ around the gimbal axis for $i\in\{1,...,N_m\}$, where $(\theta'_{Fxi},\theta'_{Fyi}) \sim\mathcal{N}(0,\sigma_{\theta_\text{aq}}^2)$ and $(\theta_{exi},\theta_{eyi})\sim\mathcal{N}(0,\sigma_{\theta_e}^2)$. Note that $\sigma_{\theta_\text{aq}}$ is a controllable search parameter that is used by the FSMs for the acquisition and $\sigma_{\theta_e}$ is the random error of FSMs, which is in the order of $\micro$rad. 
Therefore, $\theta_{Fi}$ creates a random location $p_{Fi}=(x_{Fi},y_{Fi})$ for the $i$th beam center around the $p_g$ in the $x-y$ plane) where
\begin{align}
	\left\{   \!\!\!\!\!
	\begin{array}{rl} 
		& x_{Fi} = x'_{Fi} + x_{ei},~~~ y_{Fi} = y'_{Fi} + y_{ei}, \\
		& (x'_{Fi},y'_{Fi}) \sim\mathcal{N}(0,\sigma_\text{aq}^2), ~~~
		(x_{ei},y_{ei}) \sim\mathcal{N}(0,\sigma_e^2), \\
		& \sigma_\text{aq} = Z \sigma_{\theta_\text{aq}},~~~   \sigma_e = Z \sigma_{\theta_e}.
	\end{array} \right.	\nonumber
\end{align}
The distance between the $i$th beam center and the satellite location $p_s=(x_s,y_s)$ is obtained as:
\begin{align}
	\label{a2}
	r_i &= \sqrt{(x_{Fi}-x_s)^2 + (y_{Fi}-y_s)^2} .
\end{align}
$r_i$ can also be expressed in ($x,y$) coordinate plane as:
\begin{align}
	\label{a1}
	\left\{   \!\!\!\!\!
	\begin{array}{rl} 
		& r_i = (R_{xi}+x_{ei},R_{yi}+y_{ei}),\\
		& R_i=(R_{xi},R_{yi}), ~~~~ r_{ei} = (x_{ei},y_{ei}), \\
		&R_{xi} = x'_{Fi} -x_s, ~~~~R_{yi} = y'_{Fi} -y_s.
	\end{array} \right.	
\end{align}

To get a better view, two examples of the acquisition and sensing phase are provided in Fig. \ref{sef} for $N_m=3$.  In Fig. \ref{sef1}, for sensing the location of the satellite, we see a random distribution of laser beams (the laser beam footprint of the $i$th FSM is denoted by $L_{si}$) that none of the beams are located near the satellite, and as a result, no signal is sensed by the GS. The random distribution of the beams is repeated as in Fig. \ref{sef1}, until finally, as in Fig. \ref{sef2}, a beam is placed near the satellite, and as a result, the power reflected by the satellite can be sensed by the GS. For the sensing phase, we are looking for methods that decrease the average sensing time $T_s$ while increasing the sensing accuracy. 

\subsubsection{Received Power Analysis}
The average sensing time is modeled as:
\begin{align}
	\label{se1}
	T_s = N_\text{aq} T_\text{step},
\end{align}
where alpha represents the average number of repetitions (same as Fig. \ref{sef1}) to sense the satellite (same as Fig. \ref{sef2}), $T_\text{step}= K T_\text{bit}$ is the time of each repetition/step, and $T_\text{bit}$ is the bit duration. For example, for OOK modulation with 1 Gbps data rate, $T_\text{bit}=1$ ns.
$K$ is an important parameter in sensing time. As $K$ becomes larger, the sensing accuracy improves at the cost of increasing the sensing time. The appropriate value for $K$ depends on channel parameters such as $\sigma$, $N_\text{aq}$, and channel coherence time. The coherence time of the fixed terrestrial optical channel is in the order of ms. However, due to the high speed of the LEO satellites (about 5-10 thousand m/s), the average coherence time $T_c=K_c T_\text{bit}$ is reduced to a few tens of $\micro$s. In the simulations, we show that to achieve a more accurate sense, $K=K_d K_c>> K_c$ to be able to average on the random coefficients of the channel. In other words, with time averaging, the random negative effects of the channel are reduced and we can better estimate the distance between the satellite and the laser beam center. Moreover, in the sensing phase we consider $s[k]=1$. 
Based on this averaging in the sensing phase, using \eqref{ch2}, \eqref{sb3}, and \eqref{sb4}, the received reflected signal caused by $L_{si}$ is modeled as: 
\begin{align}
	\label{d3}
	P_{rsi}  = \sum_{k=1}^K P_r[k] = R P_t 
	\underbrace{\sum_{k=1}^{K_d} \sum_{k'=1}^{K_c} h[k,k']}_{h_{si}} + \underbrace{\sum_{k=1}^K n[k]}_{n_s[K]},
\end{align}
for $i\in\{1,...,N_m\}$ and $n_s[K]\sim\mathcal{N}(0,KN_0)$, and
\begin{align}
	\label{ch4}
	&h_{si} = K_c  h_{L1}h_{L2} h_{pg} \sum_{k=1}^{K_d}  h_{ps,i}[k]   \sum_{m=1}^M   
	h_{a1}[k,m]h_{a2}[m,k],
\end{align}
where, by using \eqref{sb4}, the geometrical pointing loss of the $i$th FSM is modeled as:
\begin{align}
	\label{sb5}
	& h_{ps,i}[k]  = \frac{2A_\text{MRR}}{\pi w_{zs}^2}  \exp\left(-2\frac{(R_{xi}+x_{ei})^2 + (R_{yi}+y_{ei})^2}{w_{zs}^2} \right) .
\end{align} 
%
{\bf Theorem 1.}  {\it  For large $K_d$, the distribution of $h_{si}$ conditioned on $R_i$ is derived as:}
\begin{align}
	\label{sd1}
	f_{h_{si}|R_i}(h_{si}) \simeq \frac{1}{\sqrt{2\pi \mathbb{V}(h_{si})}} \exp\left( - \frac{(h_{si}-\mathbb{E}(h_{si}))^2}
	{2 \mathbb{V}(h_{si}) } \right),
\end{align}
{\it where $\mathbb{V}(h_{si})=\mathbb{E}(h^2_{si})-\mathbb{E}^2(h_{si})$, and }
\begin{align}
	\label{ch7}
	\mathbb{E}(h_{si}|R_i) = c_2 K_d M \mathbb{E}(h_{ps,i}|R_i) \left( \frac{\Gamma(\alpha+1) \Gamma(\beta+1)}
	{\alpha \beta \Gamma(\alpha) \Gamma(\beta)} \right)^2,
\end{align}
\begin{align}
	\label{aq6}
	&\mathbb{E}(h^2_{si}|R_i) = K_d M c_2^2 \mathbb{E}( h^2_{ps,i}|R_i) 	\Bigg[
	\left(  \frac{\Gamma(\alpha+2) \Gamma(\beta+2)}
	{\alpha^2 \beta^2 \Gamma(\alpha) \Gamma(\beta)} \right)^2  \nonumber \\
	&
	+ (M-1)  \left( \frac{\Gamma(\alpha+1) \Gamma(\beta+1)}
	{\alpha \beta \Gamma(\alpha) \Gamma(\beta)} \right)^4  \Bigg]  \\
	&+c_2^2  M^2 K_d^3(K_d-1)  \mathbb{E}^2(h_{ps,i}|R_i) \left( \frac{\Gamma(\alpha+1) \Gamma(\beta+1)}
	{\alpha \beta \Gamma(\alpha) \Gamma(\beta)} \right)^4. \nonumber
\end{align}
Also, $ c_1 = \frac{w_{zs}^2}{4\sigma_e^2}$,
$c_2=K_c  h_{L1}h_{L2} h_{pg}$, and $\mathbb{E}( h_{ps,i}|R_i)$ and $\mathbb{E}( h^2_{ps,i}|R_i)$ are obtained in \eqref{s3} and \eqref{sz1}, respectively.
\begin{IEEEproof}
	Please refer to Appendix \ref{AppA}.
\end{IEEEproof}

In the sensing phase, due to the large $R_i$ and as a result a weak SNR, it is necessary to consider more time (larger $K_d$) for each step and therefore, the analytical results of \eqref{sd1} are close to the simulations.

\subsubsection{$R_i$ Estimation}
In the sensing phase, in addition to sensing, we should be able to get an estimate of $R_i$ for use in the next phase (positioning phase). Using \eqref{sd1}, the distribution function of the $P_{rsi}$ conditioned on $R_i$ is obtained as:
\begin{align}
	\label{sd2}
	f_{P_{rsi}|R_i}(P_{rsi}) \simeq \frac{\exp\left( - \frac{(P_{rsi}-RP_t\mathbb{E}(h_{si}))^2}
		{2 (R^2 P_t^2 \mathbb{V}(h_{si}) + KN_0) } \right)}
	{\sqrt{2\pi (R^2P_t^2\mathbb{V}(h_{si})+KN_0)}} .
\end{align}
Now, using \eqref{sd2}, the ML estimate of $R_i$ is obtained as
\begin{align}
	\label{ml1}
	\hat{R}_i &= \underset{ R_i}{\text{arg max}} 
	 ~~~ f_{P_{rsi}|R_i}(P_{rsi})\nonumber \\
	&=\underset{ R_i} {\text{arg min}}  
	 ~~~  \ln(R^2P_t^2\mathbb{V}(h_{si})+KN_0) \nonumber \\
	&~~~~~~~~~~~~~  + \frac{(P_{rsi}-RP_t\mathbb{E}(h_{si}))^2}
	{R^2 P_t^2 \mathbb{V}(h_{si}) + KN_0}.
\end{align} 
The problem with estimator \eqref{ml1} is that $R_i$ is a continuous variable and has a wide range from zero to several hundreds of meters. In addition, to compute each metric, two integrals of \eqref{s3} and \eqref{sz1} must be solved, numerically. Therefore, although it has an appropriate performance, the computational load is high.

{\bf Proposition 1.}  {\it When $R_i>>\sigma_e$, the PDF of $h_{si}$ conditioned on $R_i$ is obtained in the same way as \eqref{sd1} with }
\begin{align}
	\label{h1}
	\mathbb{E}(h_{si}|R_i) \simeq \mathbb{A}_1   e^{A_2 R_i^2} 
\end{align}
\begin{align}
	\label{h2}
	&\mathbb{E}(h^2_{si}|R_i) \simeq \mathbb{A}_2  e^{A_6R_i^2 }  + \mathbb{A}_3 e^{2A_2 R_i^2}, 
\end{align}
{\it where}
\begin{align}
	\left\{
	\begin{array}{rl} 
		& \mathbb{A}_1 =  c_2 K_d M A_1 \left( \frac{\Gamma(\alpha+1) \Gamma(\beta+1)}
		{\alpha \beta \Gamma(\alpha) \Gamma(\beta)} \right)^2 , \\
		& \mathbb{A}_2 = K_d M c_2^2 A_5   	\Big[
		\left(  \frac{\Gamma(\alpha+2) \Gamma(\beta+2)}
		{\alpha^2 \beta^2 \Gamma(\alpha) \Gamma(\beta)} \right)^2   \\
		&  ~~~~~~~+ (M-1)  \left( \frac{\Gamma(\alpha+1) \Gamma(\beta+1)}
		{\alpha \beta \Gamma(\alpha) \Gamma(\beta)} \right)^4  \Big], \\
		& \mathbb{A}_3 = c_2^2  M^2 K_d^3(K_d-1)  A_1^2  \left( \frac{\Gamma(\alpha+1) \Gamma(\beta+1)}
		{\alpha \beta \Gamma(\alpha) \Gamma(\beta)} \right)^4,  \nonumber
	\end{array} \right.	
\end{align}

\begin{align}
	\left\{
	\begin{array}{rl} 
		& A_1 = \frac{2A_\text{MRR}}{\pi w_{zs}^2 \sqrt{2A_3\sigma_e^2}}, ~~~
		A_2 = \frac{2 A_4^2 \sigma_e^2 - A_3}{2 A_3 \sigma_e^2}, \\
		& A_3 = \left( \frac{2}{w_{zs}^2} + \frac{1}{2\sigma_e^2}\right), ~~A_4=\frac{-1}{2\sigma_e^2}\\
		& A_5 = \left( \frac{2A_\text{MRR}}{\pi w_{zs}^2}  \right)^2 
		\frac{ 1}{2 \sqrt{2A_7\sigma_e^2}},~~~
		A_6 = \frac{(2A_4 ^2\sigma_e^2-A_7)} {2A_7\sigma_e^2}, \\
		& A_7  = \left( \frac{4}{w_{zs}^2} + \frac{1}{2\sigma_e^2}\right). \nonumber
	\end{array} \right.	
\end{align}

\begin{IEEEproof}
	Please refer to Appendix \ref{AppB}.
\end{IEEEproof}

{\bf Proposition 2.}  {\it The PDF of $\hat{R}_i$ conditioned on $R_i$ is obtained as: }
\begin{align}
	\label{lm6}
	&f_{\hat{R}_i|R_i}(\hat{R}_i) \nonumber \\
	& = \frac{2RP_tA_2 \mathbb{A}_1 }
	{\sqrt{2\pi (R^2P_t^2\mathbb{V}(h_{si})+KN_0)}} 
	\hat{R}_i e^{A_2\hat{R}^2_i} \nonumber \\
	&~~~\times \exp\left( - 
	\frac{R^2P_t^2\mathbb{A}^2_1\left(e^{A_2\hat{R}^2_i} -e^{A_2 R_i^2}   \right)^2}
	{2 (R^2 P_t^2 \mathbb{V}(h_{si}) + KN_0) } \right),
\end{align}
{\it where}
\begin{align}
	\mathbb{V}(h_{si}) = \mathbb{A}_2  e^{A_6R_i^2 }  +  e^{2A_2 R_i^2} (\mathbb{A}_3  - \mathbb{A}^2_1  ) .
\end{align}
\begin{IEEEproof}
	Please refer to Appendix \ref{AppC}.
\end{IEEEproof}

Using Proposition 2, we can obtain the estimation error distribution based on important channel parameters such as $w_{zs}$, $\sigma_e$, and $R_i$ which helps us to optimize $w_{zs}$ to reduce the estimation errors. However, the results of Propositions 1 and 2 are valid for $R_i>>\sigma_e$ (approximately $R_i>5\sigma_e$). For example, for $\sigma_e=3$ m, the results are valid for $R_i>15$ m.  Note that according to the conventional values of $\sigma_{ge}$, which are in the order of several hundred meters to several km, the probability of the laser beam center being placed close to the satellite with $R_i<5\sigma_e$ is very low. However, if $R_i<5\sigma_e$, then the MRR array is located near the laser beam and the reflected received power is suitable enough that we can sense the satellite even with simpler techniques such as averaging to have a good estimate of $R_i$ as:
\begin{align}
	\label{g1}
	\hat{R}_i = \sqrt{ \frac{w_{zs}}{2}  \ln\left(   \frac{2K M h_L A_\text{MRR}}{   P_{rsi} \pi w_{zs}^2   } 
		\left(  \frac{\Gamma(\alpha+1) \Gamma(\beta+1)}
		{\alpha \beta \Gamma(\alpha) \Gamma(\beta)} \right)^2  \right) }.
\end{align}
The above expression is obtained by assuming large values of $K_d$ and can provide an initial estimate of $R_i$. \eqref{g1} can be used to limit the search interval of $R_i$ in the estimator \eqref{ml1}, which significantly reduces its computational load. In the simulations section, the estimation accuracy of these methods is examined.

\subsubsection{Sensing Time Analysis}
To sense the satellite, the received signal power must be greater than the threshold $P_{r\text{th}}$ as
\begin{align}
	\label{st1}
	\mathbb{S}_{\text{on}|P_{rsi}} = 
	\left\{   \!\!\!\!\!
	\begin{array}{rl} 
		& 1, ~~~~ P_{rsi}> P_{r\text{th}}, \\
		& 0, ~~~~ P_{rsi}< P_{r\text{th}}.
	\end{array} \right.	
\end{align}
According to \eqref{g1}, there is a one-to-one and inverse non-linear relationship between $P_{rsi}$ and $\hat{R}_i$ in such a way that the higher $P_{rsi}$, the smaller $\hat{R}_i$ is estimated. Therefore, \eqref{st1} can be rewritten as follows:
\begin{align}
	\label{st2}
	\mathbb{S}_{\text{on}|P_{rsi}} = 
	\left\{   \!\!\!\!\!
	\begin{array}{rl} 
		& 1, ~~~~ \hat{R}_i> R_\text{th}, \\
		& 0, ~~~~ \hat{R}_i< R_\text{th},
	\end{array} \right.	
\end{align}
where $R_\text{th}$ is the threshold sensing distance.
The sensing probability of the $i$th laser beam $L_i$ is obtained as:
\begin{align}
	\label{t1}
	\mathbb{P}_{S,\text{on},i} &= \int_0^\infty \text{Prob}\left\{ \hat{R}_i<R_\text{th} \big| R_i \right\}
	f_{R_i}(R_i) \text{d}R_i \nonumber \\
	&= \int_0^\infty   F_{\hat{R}_i}(R_\text{th}| R_i) 
	f_{R_i}(R_i) \text{d}R_i.
\end{align}

\begin{figure}
	\begin{center}
		\includegraphics[width=3.3 in]{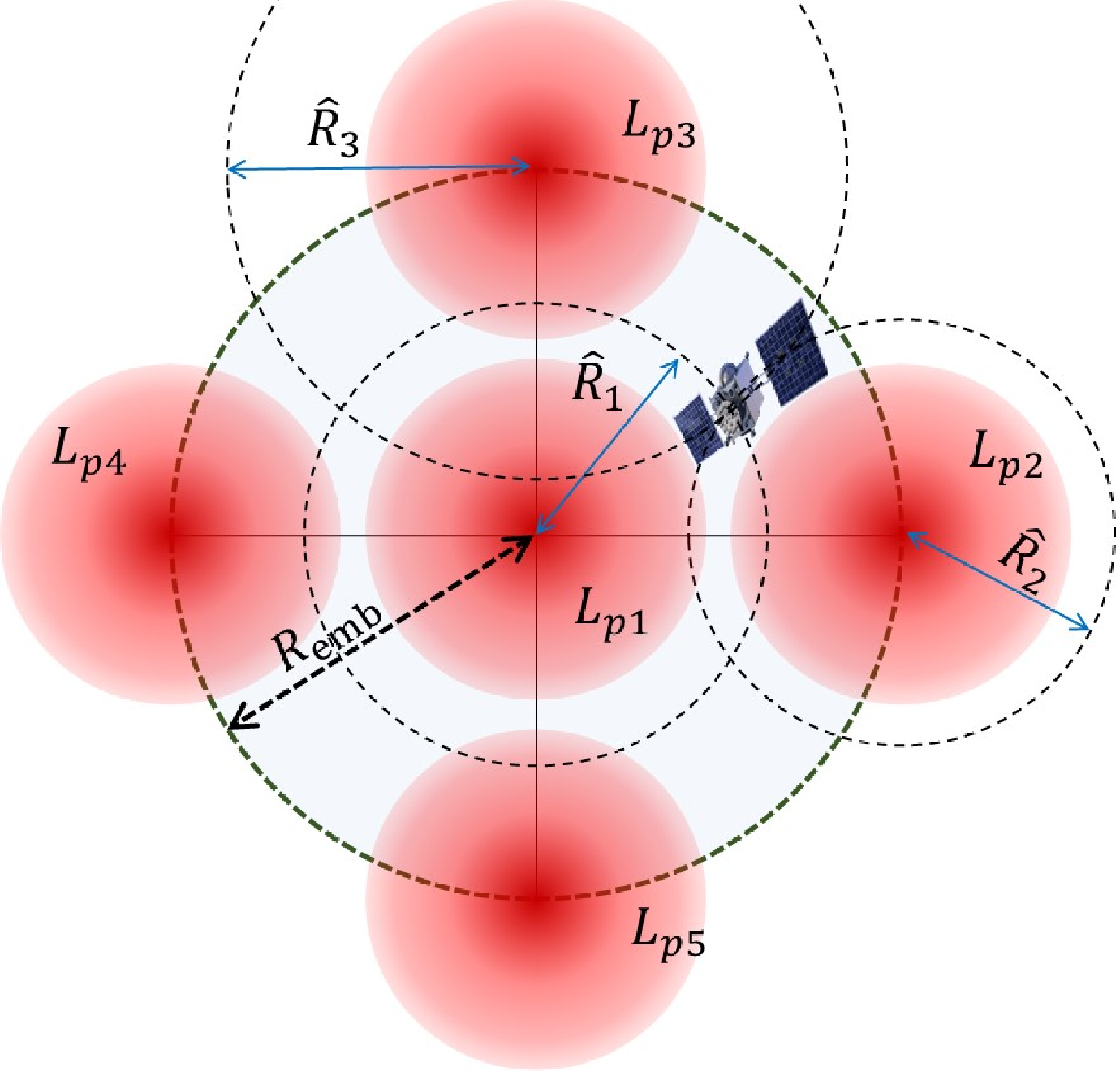}
		\caption{The considered topology features five laser beams, with one placed at the center of the ambiguity circle and the other four symmetrically positioned on the circle's circumference. In each quadrant of the ambiguity circle, the satellite is consistently covered by three laser beams.}
		\label{pos1}
	\end{center}
\end{figure}
%

As we show in the positioning phase if the estimation error $|R_i-\hat{R}_i|$ in the sensing phase is too large, then positioning is not possible and the sensing phase must be repeated. Therefore, sensing with an acceptable error can lead to satellite positioning. Let us define $R_e>|R_i-\hat{R}_i|$ as the acceptable error threshold. In this case, \eqref{t1} is modified as:
\begin{align}
	\label{t2}
	&\mathbb{P}_{S,\text{on},i} =  
	\int_0^\infty 
	\mathbb{P}_{S,\text{on},i|R_i}
	f_{R_i}(R_i) \text{d}R_i ,
\end{align}
where $\mathbb{P}_{S,\text{on},i|R_i}$ is obtained as:
\begin{align}
	\label{tt1}
	& \mathbb{P}_{S,\text{on},i|R_i} = 
	\text{Prob}\left\{ |R_i-\hat{R}_i|<R_e,\hat{R}_i<R_\text{th} \Big| R_i \right\}  \nonumber \\
	&= \left( F_{\hat{R}_i}(R_\text{th}| R_i) - F_{\hat{R}_i}(R_i-R_e| R_i)  \right) 
	\mathbb{U}(R_e-|R_i-R_\text{th}|) \nonumber \\
	& + \left( F_{\hat{R}_i}(R_i+R_e| R_i) - F_{\hat{R}_i}(R_i-R_e| R_i) \right) \nonumber \\
	&\times\left( \mathbb{U}(R_i - R_e) - \mathbb{U}(R_i-R_\text{th}+R_e) \right) \nonumber \\
	&+ F_{\hat{R}_i}(R_i+R_e| R_i) \mathbb{U}(R_e-R_i),
\end{align}
where $\mathbb{U}(x)=\left\{\!\!\!\!\!\!\!\!\!
\begin{array}{rl} 
	& 1, x>0,\\
	& 0, x<0.
\end{array} \right.	$.
Using \eqref{lm6} and \eqref{tt1}, $\mathbb{P}_{S,\text{on},i|R_i}$is derived in \eqref{tt2}, where $Q(\cdot)$ is the well-known {\it Q}-function.
\begin{figure*}[!t]
	\normalsize
	\begin{align} \label{tt2}
	&\mathbb{P}_{S,\text{on},i|R_i} =   \left[ Q\left(   \frac{RP_t \mathbb{A}_1\left( e^{A_2(R_i-R_e)^2} - e^{A_2 R_i^2}   \right)}
	{  \sqrt{R^2 P_t^2 \mathbb{V}(h_{si}) + KN_0} } \right)  -   
	Q\left(   \frac{RP_t \mathbb{A}_1\left( e^{A_2R_\text{th}^2} - e^{A_2 R_i^2}   \right)}
	{  \sqrt{R^2 P_t^2 \mathbb{V}(h_{si}) + KN_0} } \right)   \right] 
	\mathbb{U}(R_e-|R_i-R_\text{th}|) \nonumber \\
    &~~~+   \left[ Q\left(   \frac{RP_t \mathbb{A}_1\left( e^{A_2(R_i-R_e)^2} - e^{A_2 R_i^2}   \right)}
	{  \sqrt{R^2 P_t^2 \mathbb{V}(h_{si}) + KN_0} } \right)  -   
	Q\left(   \frac{RP_t \mathbb{A}_1\left( e^{A_2(R_i+R_e)^2} - e^{A_2 R_i^2}   \right)}
	{  \sqrt{R^2 P_t^2 \mathbb{V}(h_{si}) + KN_0} } \right)   \right]
	\left( \mathbb{U}(R_i - R_e) - \mathbb{U}(R_i-R_\text{th}+R_e) \right) \nonumber \\
	&~~~+ \left[ 1  -   
	Q\left(   \frac{RP_t \mathbb{A}_1\left( e^{A_2(R_i+R_e)^2} - e^{A_2 R_i^2}   \right)}
	{  \sqrt{R^2 P_t^2 \mathbb{V}(h_{si}) + KN_0} } \right)   \right]
	F_{\hat{R}_i}(R_i+R_e| R_i) \mathbb{U}(R_e-R_i). 
	\end{align}
	\hrulefill
\end{figure*}  

Since the sensing phase can be performed by different laser beams $L_{si}$, the probability of sensing by a set of laser beams $N_m$ is obtained as:
\begin{align}
	\label{t4}
	\mathbb{P}_{s,\text{on}}  &= 1- \mathbb{P}_{s,\text{off}} 
	= 1- \prod_{i=1}^{N_m} \mathbb{P}_{s,\text{off},i} \nonumber \\
	&  = 1- \prod_{i=1}^{N_m} (1-\mathbb{P}_{s,\text{on},i} )
	= 1 - (1 - \mathbb{P}_{s,\text{on},i})^{N_m},
\end{align}
where $\mathbb{P}_{s,\text{off},i}$ is the probability that the satellite is sensed by the $L_{si}$.
Note that $\mathbb{P}_{s,\text{on}}$ is the sensing probability for one step. For sensing, we perform acquisition in different time steps to finally sense the satellite. According to \eqref{se1}, $N_\text{aq}$ represents the number of acquisition steps, and $T_\text{step}=K_dK_cT_\text{bit}$ is the time allocated to each step. Therefore, the average time for sensing the satellite is obtained as:
\begin{align}
	\label{tt3}
	\bar{T}_s = T_\text{step} \bar{N}_\text{aq} = {T_\text{step}}/{\mathbb{P}_{s,\text{on}}},
\end{align}
where $\bar{N}_\text{aq} =1/\mathbb{P}_{s,\text{on}}$ is the average number of steps.
For example, if $T_\text{bit}=1$ ns, $K_c=1000$, $K_d=1000$, and $\mathbb{P}_{s,\text{on}}=0.01$, then the average number of steps for sensing the satellite $\bar{N}_\text{aq}=1/\mathbb{P}_{s,\text{on}}=100$ and the average sensing time is $\bar{T}_s=K_d K_c \bar{N}_\text{aq} T_\text{bit} = 0.1$ s.

In summary, in this section, the sensing phase was investigated, which is a function of parameters $R_e$, $K_d$, $R_\text{th}$, $w_{zs}$, $\sigma_e$, and $\sigma_{ge}$. In the simulations, the effect of these parameters on the accuracy and sensing time is investigated.

\subsection{Positioning Phase}
In the sensing phase, in addition to satellite sensing, an initial estimate of the distance of the satellite relative to the $i$th reflected laser beam is made, which can have a tolerable error as high as $R_e$. Therefore, after the sensing phase, we know that the satellite is most likely in a circle with radius $R_\text{emb}=\hat{R}_i+R_e$ around the center of the $i$th laser beam. We call this circle the ambiguity circle. In the positioning phase, the target is to obtain accurate estimation of the location of the satellite in the ambiguity circle. Unlike distance, in the positioning phase, we need at least three laser beams to estimate location \cite{cannizzaro2019comparison}. Based on this, we propose a topology based on five laser beams as we set one laser beam in the center of the ambiguity circle, and the other four beams are set symmetrically on the circumference of the ambiguity circle as shown in Fig. \ref{pos1}. In this proposed topology, depending on the location of the satellite in each quadrant of the ambiguity circle, it is always covered by three laser beams. Similar to the sensing phase, the reflected signal received by the $i$th laser beam is modeled as:
\begin{align}
	\label{p1}
	P_{rpi}  = \sum_{k=1}^{K_d} P'_{rpi}[k] = 
	\sum_{k=1}^{K_d}  (R P_t K_c h[k] +  n_p[k]),
\end{align}
where $n_p[k]\sim\mathcal{N}(0,K_cN_0)$.
Similar to the sensing phase, we represents the distance between the center of the $i$th laser beam (denoted by $L_{pi}$) and the satellite with $r_i = (R_{xi}+x_{ei},R_{yi}+y_{ei})$, where $ R_i=(R_{xi},R_{yi})$, $R_{xi} = x_{pi} -x_s$, $R_{yi} = y_{pi} -y_s$, $p_{pi}=(x_{pi},y_{pi})$ is the average location of $L_{pi}$, and $p_s=(x_s,y_s)$ is the location of the satellite inside the ambiguity circle. According to Fig. \ref{pos1}, without loss of generality, we assume that 
$p_{p1}=(0,0)$, $p_{p2}=(R_\text{emb},0)$, $p_{p3}=(0,R_\text{emb})$, $p_{p4}=(-R_\text{emb},0)$, and $p_{p5}=(0,-R_\text{emb})$. 
Also, similar to the sensing phase, $(x_{ei},y_{ei})\sim\mathcal{N}(0,\sigma_e^2)$ are the spatial pointing errors caused by FSMs and atmospheric turbulence induced beam wandering.
Moreover, due to the symmetry of the problem, without loss of generality, we assume that the satellite is always in the quarter of the circle that is between $L_{p1}$, $L_{p2}$, and $L_{p3}$. Using $R_1$, $R_2$, and $R_3$, we obtain the location of the satellite as:
\begin{align} \label{p2}
	\left\{   \!\!\!\!\!\!
	\begin{array}{rl} 
		& x_s = \frac{R_1^2-R_2^2 + R_\text{emb}^2}{2 R_\text{emb}}, \\
		& y_s = \frac{R_1^2-R_3^2 + R_\text{emb}^2}{2 R_\text{emb}}.
	\end{array} \right.	
\end{align}
To use \eqref{p2}, the values of $R_1$, $R_2$, and $R_3$ are not known and must be estimated. Unlike the sensing phase, here the values of $R_1$, $R_2$, and $R_3$ are small (in the order of usually less than 100 m) and thus, it is better to use the proposed method in \eqref{g1} to estimate $R_i$s. However, here, the estimation method in \eqref{g1}  can be performed in two ways. The first method is to estimate $\hat{R}_i$s based on the received $P_{rpi}$. The second method is to estimate $\hat{R}'_i[k]$ based on $P'_{rpi}[k]$, and finally estimate $\hat{R}_i$ by averaging as:
\begin{align}
	\label{p3}
	\hat{R}_i = \frac{1}{K_d}\sum_{k=1}^{K_d} \hat{R}'_i[k].
\end{align}
Then, substituting the estimated $\hat{R}_i$ instead of $R_i$s in \eqref{p2}, the location of the satellite is estimated.
In the simulations section, it is shown that the performance of both methods depends on $w_z$, and the advantages and disadvantages of both methods are examined. 

For comparison, we compare the performance of both positioning methods with the ideal estimator. For the ideal estimator, we assume that the receiver noise is zero and the atmospheric fluctuations are constant and equal to its average value. In this case, the changes in the received power are only caused by $r_i$. Therefore, for ideal estimator, we obtain:
\begin{align} \label{p4}
	\left\{   \!\!\!\!\!\!
	\begin{array}{rl} 
		& \hat{x}_{s,\text{ideal}}  = \frac{\sum_{k=1}^{K_d}\left(  (R_{x1}+x_{e1}[k])^2+(R_{y1}+y_{e1}[k])^2 \right)}
		{{2 K_d R_\text{emb}}}  \\
		& ~~~~~- \frac{\sum_{k=1}^{K_d}\left(  (R_{x2}+x_{e2}[k])^2+(R_{y2}+y_{e2}[k])^2 \right)}
		{{2 K_d R_\text{emb}}} + \frac{R_\text{emb}}{2}, \\
		& \hat{y}_{s,\text{ideal}}  = \frac{\sum_{k=1}^{K_d}\left(  (R_{x1}+x_{e1}[k])^2+(R_{y1}+y_{e1}[k])^2 \right)}
		{{2 K_d R_\text{emb}}}  \\
		& ~~~~~- \frac{\sum_{k=1}^{K_d}\left(  (R_{x3}+x_{e3}[k])^2+(R_{y3}+y_{e3}[k])^2 \right)}
		{{2 K_d R_\text{emb}}} + \frac{R_\text{emb}}{2}.
	\end{array} \right.	
\end{align}
Note that \eqref{p4} is only used as a benchmark to compare the performance of the positioning methods.

Finally, note that the absolute value of the positioning error causes the geometrical pointing loss, which is obtained as follows:
\begin{align}
	\label{ps1}
	R_\text{ep} = \sqrt{(\hat{x}_s-x_s)^2 + (\hat{y}_s-y_s)^2}.
\end{align}
In the simulations, the metric $R_\text{ep}$ is used to evaluate the effect of different system parameters on the performance of the positioning phase.

\begin{table}
	\caption{Parameter values used in the simulation.} 
	\centering 
	\begin{tabular}{c } 
		\includegraphics[width=3 in]{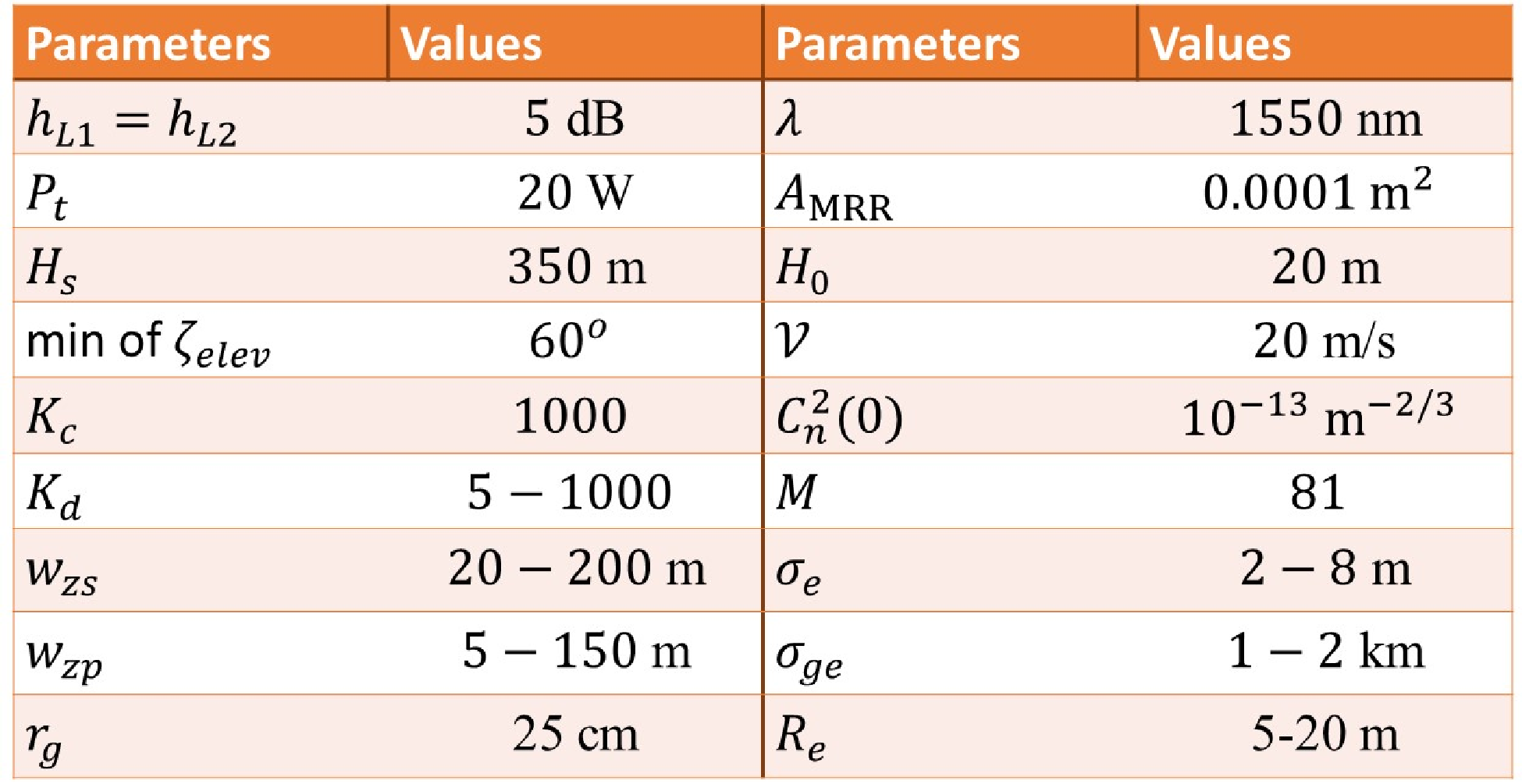}     
	\end{tabular}
	\label{Tab1} 
\end{table}

\section{Simulation Results and Discussion}
In this section, by using simulations, the performance of the considered system is examined in both the sensing phase and the positioning phase. The values of the simulation parameters are provided in Table \ref{Tab1}, which are based on the typical values available in the literature \cite{andrews2005laser,kaushal2016optical}.

\subsection{Sensing Phase}
The sensing phase consists of a random search in the ambiguous area using several laser beams. This random search is repeated in different time steps until finally a laser beam is located at a short distance from the satellite. In this case, the signal reflected from the MRR array can be sensed by the GS. After that, using the received reflected signal,  the GS estimates $R_i$. Sensing accuracy depends on the estimation accuracy of $R_i$. 

\begin{figure}
	\centering
	\subfloat[] {\includegraphics[width=3.2 in]{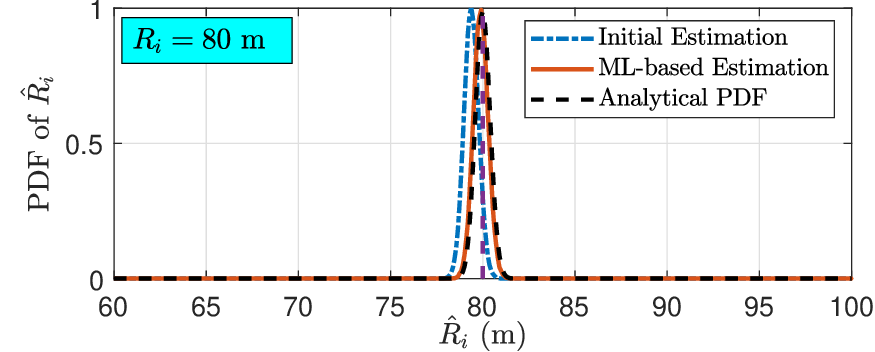}
		\label{sf1}
	}
	\hfill
	\subfloat[] {\includegraphics[ width=3.2 in]{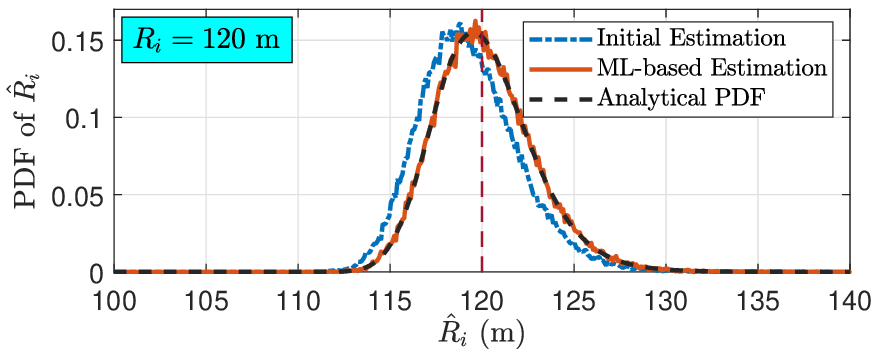}
		\label{sf2}
	}
	\hfill
	\subfloat[] {\includegraphics[width=3.2 in]{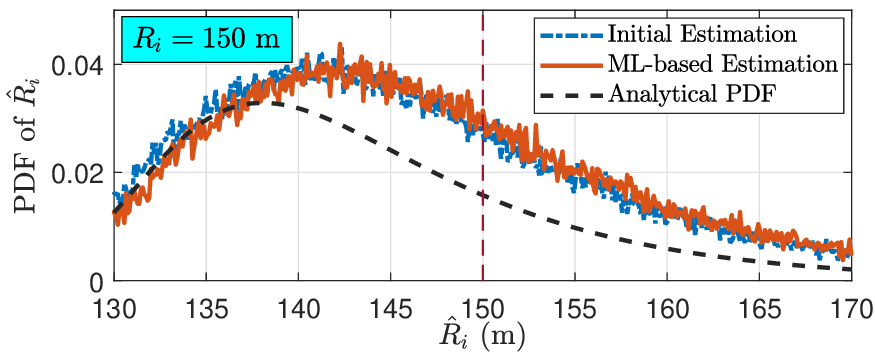}
		\label{sf3}
	}
	\caption{The PDF of $\hat{R}_i$ is provided for $w_{zs}=80$, $K_d=500$, and  three different values of (a) $R_i=80$ m; (b) $R_i=120$ m; and (c) $R_i=150$ m.  
	}
	\label{sf}
\end{figure}

To evaluate the accuracy of the proposed methods, in Fig. \ref{sf}, the PDF of $\hat{R}_i$ is provided for three different values of $R_i=80$, 120, and 150 m. The results of Fig. \ref{sf} are obtained for $w_{zs}=80$ and $K_d=500$. By increasing $R_i$, the geometrical pointing loss increases, and thus, the SNR decreases. Therefore, as we expect, the results of Fig. \ref{sf} show that with the increase of $R_i$, the variance of the estimation errors increases and the accuracy of the sensing phase decreases. In other words, the results show the relationship between sensing time and sensing accuracy. If a large $R_\text{th}$ is chosen for the sensing metric, the sensing time reduces at the cost of lower accuracy for $\hat{R}_i$.

\begin{figure}
	\centering
	\subfloat[] {\includegraphics[width=3.2 in]{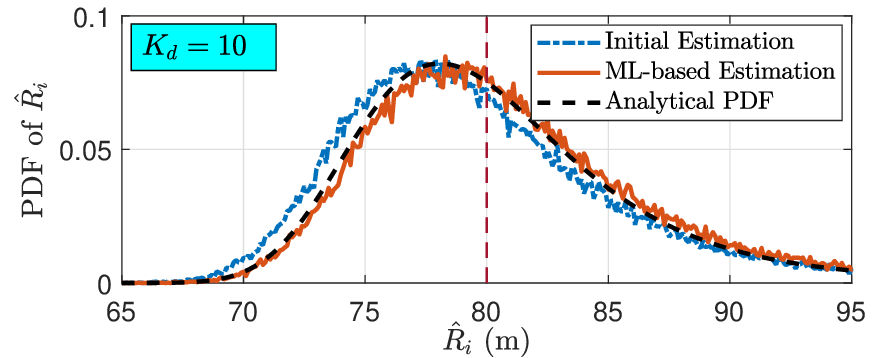}
		\label{sm1}
	}
	\hfill
	\subfloat[] {\includegraphics[ width=3.2 in]{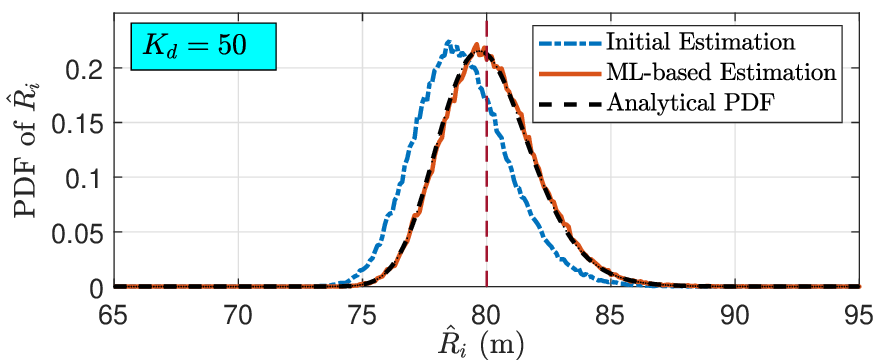}
		\label{sm2}
	}
	\hfill
	\subfloat[] {\includegraphics[width=3.2 in]{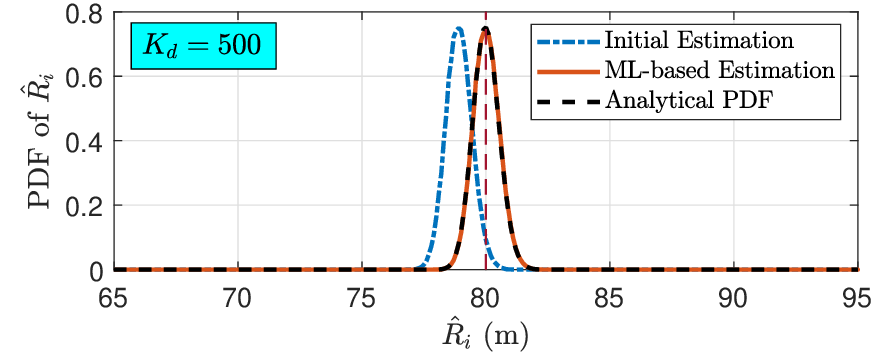}
		\label{sm3}
	}
	\caption{Investigating the effect of $K_d$ on the accuracy of $\hat{R}_i$ for (a) $K_d=10$; (b) $K_d=50$; and (c) $K_d=500$. 
	}
	\label{sm}
\end{figure}

In Fig. \ref{sm}, the effect of $K_d$ on the accuracy of $\hat{R}_i$ is investigated.
The results of Figs. \ref{sm1}-\ref{sm3} are plotted for $K_d=10$, 50, and 500, respectively. The results show that with the increase of $K_d$, the variance of the estimation errors decreases. However, since there is a linear relationship between the sensing time and $K_d$, this increase in accuracy again comes at the cost of more sensing time. In addition, by comparing both provided estimation methods, it can be seen that the ML-based estimator has a better performance, and although the initial estimation method (averaging-based method) provided in \eqref{g1} has a lower computational load, it causes an estimation bias error. Also, the results of both Figs. \ref{sf} and \ref{sm} show that the derived analytical expression (provided in Proposition 2) can accurately model the PDF of $\hat{R}_i$ and can be used for further analysis.

\begin{figure}
	\centering
	\subfloat[] {\includegraphics[width=3.3 in]{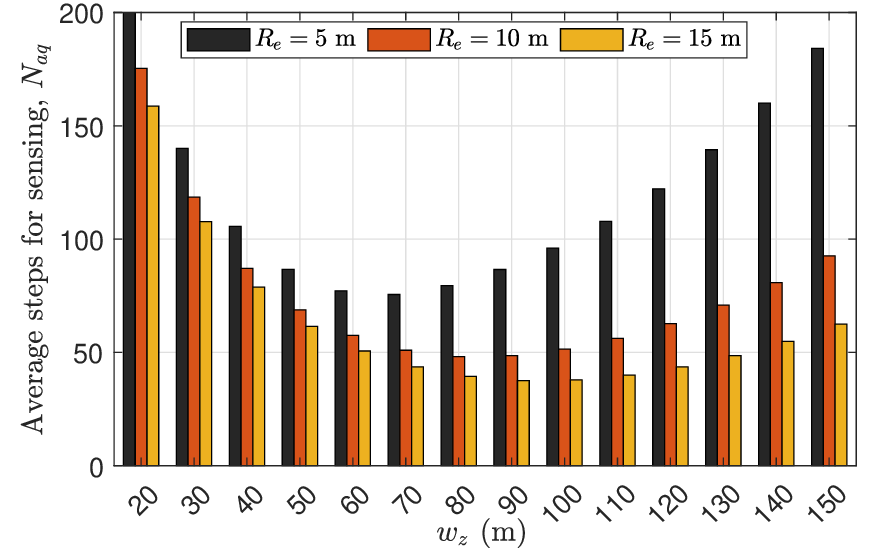}
		\label{sft1}
	}
	\hfill
	\subfloat[] {\includegraphics[ width=3.3 in]{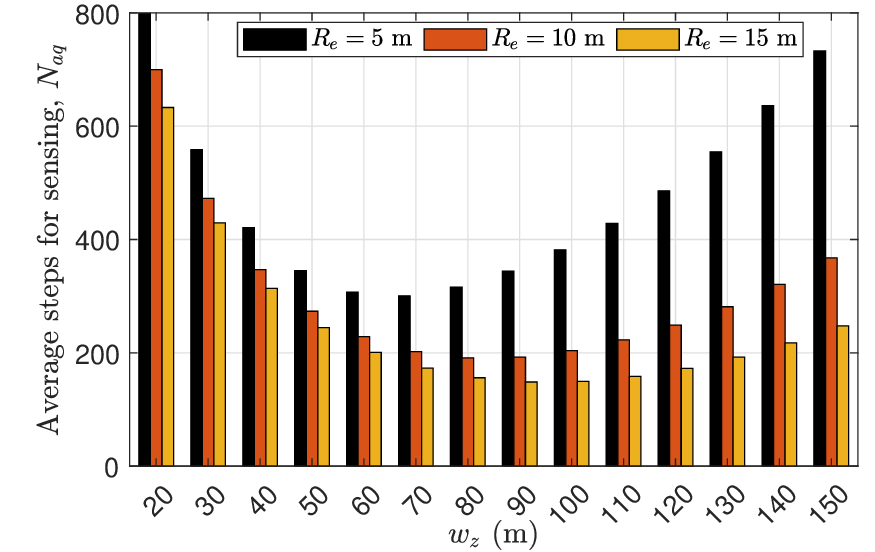}
		\label{sft2}
	}
	\caption{The sensing time (characterized by $N_\text{aq}$) versus $w_{zs}$  for different values of $R_e$ and (a) $\sigma_{ge}=1$ km; and (b) $\sigma_{ge}=2$ km.  
	}
	\label{sft}
\end{figure}

Unlike conventional optical-based satellite communications, for the considered MRR-based optical satellite system, $w_{zs}$ is a very important parameter and plays a decisive role in the quality and speed of the sensing phase. To this end, the sensing time (characterized by $N_\text{aq}$) is plotted versus $w_{zs}$ in Figs. \ref{sft1} and \ref{sft2} for $\sigma_{ge}=1$ and 2 km. The results are obtained for $R_i=150$ m and three different values of the sensing accuracy characterized by $R_e=5$, 10, and 15 m. The results clearly show the importance of finding the optimal $w_{zs}$ to reduce the sensing time. For example, in Fig. \ref{sft1}, for $\sigma_{ge}=1$ km, the optimal values of $w_{zs}$ to estimate $R_i$ with an error of less than $R_e=5$ and 15 m are equal to 70 and 90 m, respectively. Note that $N_\text{aq}$ is the average number of steps to sense the satellite and has a direct linear relationship with the sensing time. Also, based on the results of Fig. \ref{sft2}, with the increase of the $\sigma_{ge}$ from 1 km to 2 km (this error is caused by the tracking error of the mechanical gimbal system), the average number of steps to sense the satellite increases. For example, for satellite sensing with $R_e=10$ m, with the increase of $\sigma_{ge}$ from 1 km to 2 km, $N_\text{aq}$ increases from 46 to 194, which almost triples the sensing time.

\subsection{Positioning Phase}

\begin{figure}
	\centering
	\subfloat[] {\includegraphics[width=3.2 in]{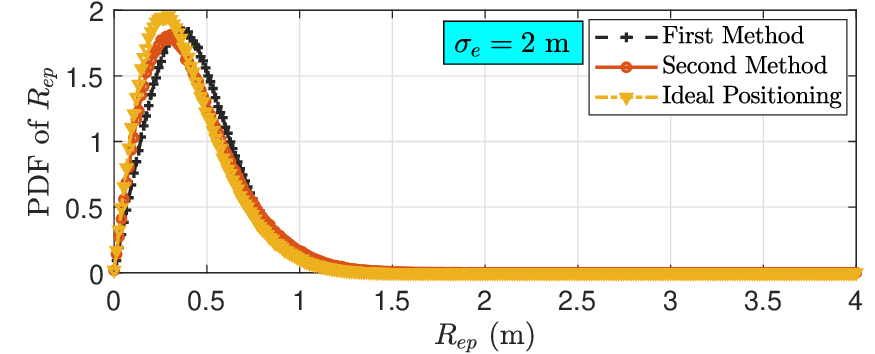}
		\label{pm1}
	}
	\hfill
	\subfloat[] {\includegraphics[ width=3.2 in]{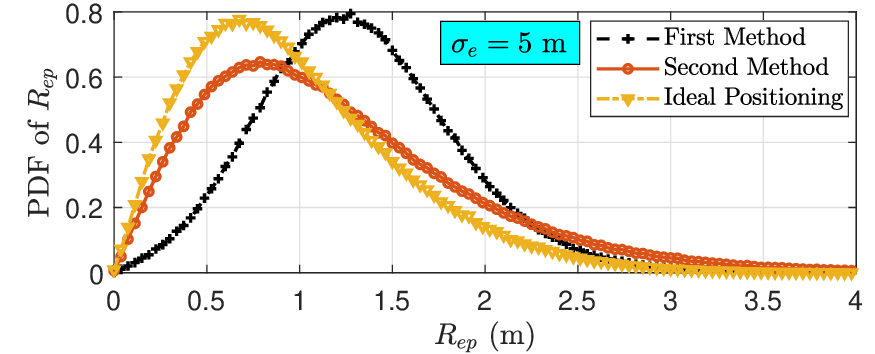}
		\label{pm2}
	}
	\hfill
	\subfloat[] {\includegraphics[width=3.2 in]{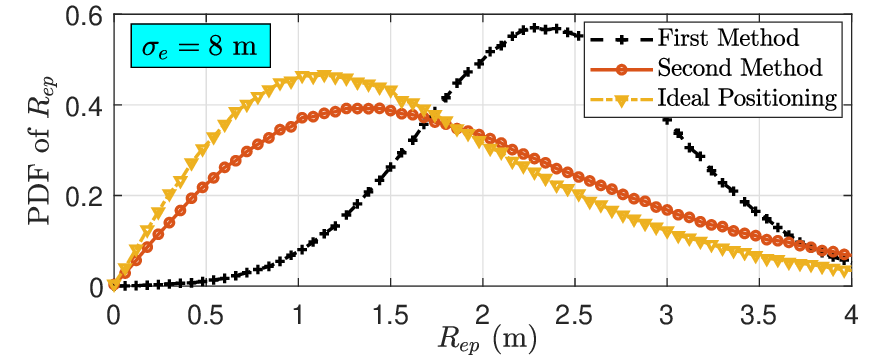}
		\label{pm3}
	}
	\caption{Evaluating the effect of $\sigma_e$ on the positioning errors for (a) $K_d=50$, $R_\text{emb}=30$ m, and (a) $\sigma_e=2$ m; (b) $\sigma_e=5$ m; and (c) $\sigma_e=8$ m. The positioning errors caused by two positioning methods are compared with the ideal positioning obtained based on \eqref{p4}.
	}
	\label{pm}
\end{figure}

After sensing the reflected signal from the satellite and estimating the initial value of $R_i$, we enter the positioning phase. The target of the positioning phase is to estimate a more accurate location of the satellite relative to the center of $L_{p1}$ as shown in Fig. \ref{pos1}. Two parameters, $\hat{R}_i$ and $\sigma_e$, affect the accuracy of the location estimation in the positioning phase. $\sigma_e$ is the beam center displacements due to the MRR errors and atmospheric turbulence-induced beam wandering. In Fig. \ref{pm}, we first evaluate the effect of $\sigma_e$ on the positioning errors for $K_d=50$ and $R_\text{emb}=30$ m. In this figure, the positioning errors caused by two positioning methods are compared with the ideal positioning obtained based on \eqref{p4}. Note that both positioning methods have the same computational complexity. As we see, for $\sigma_e=2$ m, the positioning errors of both methods are almost equal and close to the ideal positioning. It can be also seen that with the increase of $\sigma_e$ from 2 to 8 m, the positioning error increases, and the performance of the second method outperforms the first positioning method. Notice that the results of Fig. \ref{pm} are for a small value of $R_\text{emb}=30$ m.

\begin{figure}
	\begin{center}
		\includegraphics[width=3.3 in]{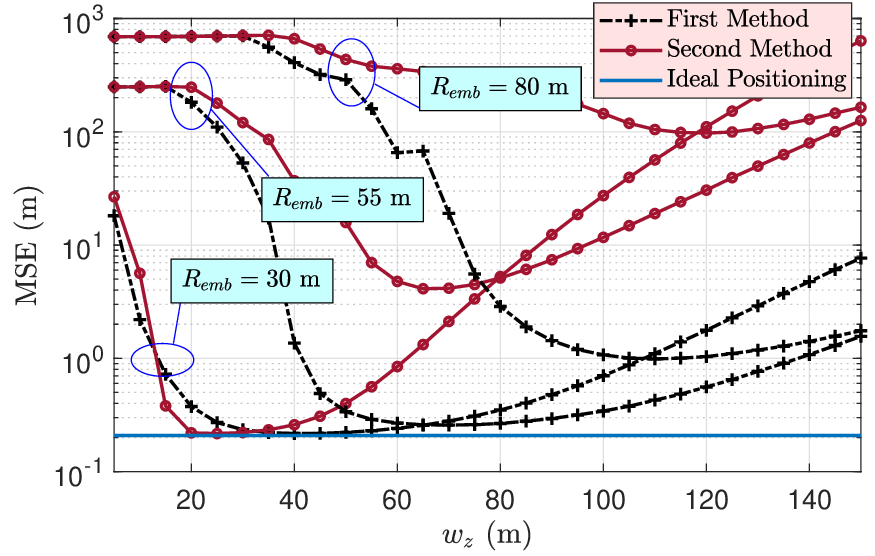}
		\caption{MSE=$\mathbb{E}(|R_{ep}|^2)$ of positioning error versus $w_{zp}$ for three different values of (a) $R_\text{emb}=30$ m; (b) $R_\text{emb}=55$ m; and (c) $R_\text{emb}=80$ m.}
		\label{pfb1}
	\end{center}
\end{figure}
%

 In Fig. \ref{pfb1}, the effect of $R_\text{emb}$ on the positioning error is depicted, where MSE=$\mathbb{E}(|R_{ep}|^2)$ is plotted versus $w_{zp}$ for three different values of $R_\text{emb}=30$, 55, and 80 m. Note that $R_\text{emb}=\hat{R}_i+R_e$ and it depends on the value of $\hat{R}_i$ estimated in the sensing phase. It can be observed that as the MSE increases by increasing $R_\text{emb}$. Unlike the results of Fig. \ref{pm}, here, as $R_\text{emb}$ increases, the first method outperforms the second positioning method. 
 As a result, for lower values of $R_\text{emb}$, it is better to use the second positioning method, and for larger values of $R_\text{emb}$, the first method has lower errors. Another important point is that the performance of both estimators is a function of  $w_{zp}$, and for each $R_\text{emb}$, the optimal value of $w_{zp}$ is different. Therefore, the results indicate that during the implementation of such an MRR-based system, the laser transmitter of the GS must have the ability to change the divergence angle and, as a result, select the optimal $w_{zp}$ according to each instantaneous value of $R_\text{emb}$.

\section{Concluding Remarks and Future Directions}
In this paper, we have focused on the optimal design of an MRR-based laser link to establish a high-speed downlink for CubeSats, addressing the weight and power limitations commonly encountered by these tiny satellites. We have demonstrated the importance of precise alignment accuracy for this round-trip downlink with high geometrical pointing loss, directing our attention toward the analysis and design of sensing and positioning systems. Moreover, the interplay between system parameters, such as mechanical gimbal error and FSM alignment error has been explored along with their impacts on the sensing time and the positioning accuracy. Finally, we have illustrated that achieving high accuracy, which is a prerequisite for establishing a high-rate link, necessitates the GS's capability to optimize and select the optimal divergence angle, effectively controlling the laser beamwidth under various momentary conditions.

After the positioning phase, we have achieved sufficient alignment accuracy to enter the communication phase. In the communication phase, the OOK pulse duration is in the order of ns (whereas it is $K=K_cK_d$ times lower than the received signal time interval for the sensing and positioning phase), and therefore to increase the SNR, it is necessary to reduce the laser beamwidth to increase the power of the reflected signal while maintaining high alignment accuracy. On the other hand, it makes the system sensitive to alignment errors. Therefore, due to the high speed of the LEO satellite, in the communication phase, the tracking phase should be used at the same time, which can be an attractive topic for future work of the considered asymmetric MRR-based satellite system.

\appendices

\section{}\label{AppA}

Using \eqref{sb5}, the CDF of $h_{ps,i}$ conditioned on $R_i$ is obtained as follows:
\begin{align}
	\label{sbr}
	&F_{h_{ps,i}|R_i}(h_{ps,i}) =
	\text{Prob}\Bigg[  r_i > 
	\sqrt{\frac{w_{zs}^2}{2}     \ln\left( \frac{\pi w_{zs}^2 h_{ps,i}} {2A_\text{MRR}}  \right)}  \Big|    R_i    \Bigg] \nonumber \\
	&~~~~~~ = 1 -  F_{r_i|R_i}\left( \sqrt{\frac{w_{zs}^2}{2}     \ln\left( \frac{\pi w_{zs}^2 h_{ps,i}} {2A_\text{MRR}}  \right)}   \right),
\end{align}
Using \eqref{a2} and \eqref{a1}, the distribution of $r_i$ conditioned on $R_i$ is derived as \cite{Raician1}:
\begin{align}
	\label{s1}
	&f_{r_i|R_i}(r_i) = \frac{r}{\sigma_e^2} \exp\left( -\frac{r_i^2+R_i^2}{2\sigma_e^2} \right)
	I_0\left( \frac{r_i R_i}{\sigma_e^2}  \right) . 
\end{align}
Using \eqref{sbr} and \eqref{s1}, the PDF of $h_{ps,i}$ conditioned on $R_i$ is obtained as:
\begin{align}
	\label{s2}
	& f_{h_{ps,i}|R_i}(h_{ps,i}) =\! f_{r_i|R_i}\left( \sqrt{\frac{w_{zs}^2}{2} \!    \ln\left( \frac{\pi w_{zs}^2 h_{ps,i}} {2A_\text{MRR}}  \right)}   \right)\!\times\! \frac{\text{d} r_i}{ \text{d} h_{ps,i} } \nonumber \\
	&  =  \frac{c_1   e^{ -\frac{R_i^2}{2\sigma_e^2} }  }
	{ \left( \frac{2A_\text{MRR}}{\pi w_{zs}^2} \right)^{c_1}  }  
	h_{ps,i}^{c_1-1}
	I_0\left( \frac{ R_i w_{zs}\sqrt{\ln\left(\frac{2A_\text{MRR}}{\pi w_{zs}^2h_{ps,i}} \right)} }{\sqrt{2}\sigma_e^2}  \right) ,
\end{align}
where $  0<h_{ps,i}<\left(\frac{2A_\text{MRR}}{\pi w_{zs}^2}\right)$, $ c_1 = \frac{w_{zs}^2}{4\sigma_e^2}$, and $I_0(\cdot)$ is the modified Bessel function of the first kind with order zero. Now, we define $h_{aM} = \sum_{m=1}^M  h_{a1}[k,m]h_{a2}[m,k]$. Here, we assume that $h_{a1}[k,m]$s and $h_{as}[k,m]$s are independent for $m\in\{1,...,M\}$ \cite{10299800}. 
To estimate $R_i$ using the ML method, we need the PDF  of $h_{si}$.
According to \eqref{ch4}, $h_{si}$ consists of the multiplication and addition of a large number of independent random variables, which are complicated to obtain an analytical closed-form expression for the PDF of $h_{si}$. 
According to the central limit theorem, the sum of large independent random variables tends to the Gaussian distribution function. Therefore, for large $K_d$, the PDF of $h_{si}$ can be well modeled by the Gaussian distribution in \eqref{sd1}, where its parameters are obtained below. It is necessary to note that the MRR array reduces the intensity of atmospheric fluctuations by creating spatial diversity. In addition, in the sensing phase, due to the large average values of $R_i$, a large $w_{zs}$ is selected (in the order of tens of meters to hundreds of meters), which reduces the variance of the geometrical pointing loss. Therefore, it can be seen that the Gaussian approximation for $h_{si}$ is accurate even for $K_d>5$.
Using \eqref{ch4}, the expected value of $h_{si}$ conditioned on $R_i$ is obtained as
\begin{align}
	\label{ch6}
	\mathbb{E}(h_{si}|R_i) &=\! c_2 \sum_{k=1}^{K_d}\! \! \mathbb{E}(h_{ps,i}[k]|R_i)  \! \sum_{m=1}^M \!\!  
	\mathbb{E}(h_{a1}[k,m]h_{a2}[m,k]) \nonumber \\
	& = c_2 K_d M \mathbb{E}(h_{ps,i}|R_i) \mathbb{E}(h_{a1})\mathbb{E}(h_{a2})
\end{align}
where $c_2=K_c  h_{L1}h_{L2} h_{pg}$. 
Using \eqref{s2}, we have 
\begin{align}
		\label{s3}
		& \mathbb{E}(h_{ps,i}|R_i) =  \frac{c_1   e^{ -\frac{R_i^2}{2\sigma_e^2} }  }
		{ \left( \frac{2A_\text{MRR}}{\pi w_{zs}^2} \right)^{c_1}  } \int_0^{\frac{2A_\text{MRR}}{\pi w_{zs}^2}}   
		h_{ps,i}^{c_1}\nonumber \\
		& \times 
		I_0\left( { R_i w_{zs}\sqrt{\ln\left(\frac{2A_\text{MRR}}{\pi w_{zs}^2h_{ps,i}} \right)} }
		\Bigg/{\sqrt{2}\sigma_e^2}  \right) \text{d} h_{ps,i}.
\end{align}
Any $h_{a1}$ variable that has a GG distribution can be written as a product of two $h_{1a1}$ and $h_{2a1}$ variables as $h_{a1}=h_{1a1}h_{2a1}$ \cite{karagiannidis2006bounds}, where
\begin{align}
	\label{as3}
	\left\{
	\begin{array}{rl} 
		&f_{h_{1a1}}(h_{1a1})= \frac{\alpha^\alpha h_{1a1}^{\alpha-1}}{\Gamma(\alpha)}e^{-\alpha h_{1a1}}, \\
		&f_{h_{2a1}}(h_{2a1})= \frac{\beta^\beta h_{2a1}^{\beta-1}}{\Gamma(\beta)}e^{-\beta h_{2a1}}.
	\end{array} \right.	
\end{align}
Similarly, we can rewrite $h_{a2}=h_{1a2}h_{2a2}$. Based on this and using \cite{Gam_func1}, $\mathbb{E}(h_{a1})=\mathbb{E}(h_{a2})=\mathbb{E}(h_a)$ is obtained as
\begin{align}
	\label{d6}
	&\mathbb{E}(h_{a1}) = \mathbb{E}(h_{1a1}h_{2a1}) = \mathbb{E}(h_{1a1})\mathbb{E}(h_{2a1}) \nonumber \\
	& = \int_0^\infty \frac{\alpha^\alpha h_{1a1}^{\alpha}}{\Gamma(\alpha)}e^{-\alpha h_{1a1}} \text{d} h_{1a1}
	\int_0^\infty \frac{\beta^\beta h_{2a1}^{\beta}}{\Gamma(\beta)}e^{-\beta h_{2a1}} \text{d} h_{2a1} \nonumber \\
	& = \frac{\Gamma(\alpha+1) \Gamma(\beta+1)}
	{\alpha \beta \Gamma(\alpha) \Gamma(\beta)}.
\end{align}
Substituting \eqref{s3} and \eqref{d6} in \eqref{ch6}, $\mathbb{E}(h_{si}|R_i)$ is derived in \eqref{ch7}.

Now, using \eqref{ch4}, the expected value of $h^2_{si}$ conditioned on $R_i$ is obtained as
\begin{align}
	\label{aq1}
	\mathbb{E}(h^2_{si}|R_i) &= c_2^2 \mathbb{E}\Bigg( \bigg( \sum_{k=1}^{K_d} h_{pa}[k] \bigg)^2 \bigg| R_i \Bigg) ,
\end{align}
where
\begin{align}
	\label{aq2}
	h_{pa}[k] = h_{ps,i}[k]   \sum_{m=1}^M  h_{a1}[k,m]h_{a2}[m,k].
\end{align}
Due to the independence of $h_{pa}[k]$s from each other, \eqref{aq1} is simplified as
\begin{align}
	\label{aq3}
	&\mathbb{E}(h^2_{si}|R_i) = c_2^2 \mathbb{E}\Bigg(  \sum_{k=1}^{K_d} h^2_{pa}[k] \bigg|R_i \Bigg)  \nonumber \\
	&~~~+c_2^2 \mathbb{E}\Bigg(  \sum_{k=1}^{K_d} \sum_{\substack{k'=1 \\k'\neq k}}^{K_d} h_{pa}[k] h_{pa}[k']  \bigg|R_i \Bigg) \nonumber \\
	& = K_d c_2^2 \mathbb{E}( h^2_{pa}|R_i )  
	+K_d(K_d-1)\mathbb{E}^2(h_{si} |R_i),
\end{align}
where
\begin{align}
	\label{aq4}
	&\mathbb{E}( h^2_{pa}[k]|R_i ) =  \mathbb{E}\left( \left(h_{ps,i}   \sum_{m=1}^M  h_{a1}[m]h_{a2}[m]\right)^2 
	\bigg| R_i\right) \nonumber \\
	& = \mathbb{E}( h^2_{ps,i}|R_i) \Bigg[ \mathbb{E}\left( \sum_{m=1}^M  h_{a1}^2[m]h_{a2}^2[m] \right) \nonumber \\
	&~~~+ \mathbb{E}\Bigg( \sum_{m=1}^{M} \sum_{\substack{m'=1 \\m'\neq m}}^{M}  h_{a1}[m]h_{a2}[m]   h_{a1}[m']h_{a2}[m']  \Bigg)  \Bigg] \nonumber \\
	& = \mathbb{E}( h^2_{ps,i}|R_i) \Big[ M\mathbb{E}^2( h_{a1}^2) 
	+ M(M-1)  \mathbb{E}^4( h_{a1})  \Big].
\end{align}
Using \eqref{s2}, we have  
	\begin{align}
		\label{sz1}
		& \mathbb{E}(h_{ps,i}|R_i) =  \frac{c_1   e^{ -\frac{R_i^2}{2\sigma_e^2} }  }
		{ \left( \frac{2A_\text{MRR}}{\pi w_{zs}^2} \right)^{c_1}  } \int_0^{\frac{2A_\text{MRR}}{\pi w_{zs}^2}}   
		h_{ps,i}^{c_1+1}\nonumber \\
		& \times 
		I_0\left( { R_i w_{zs}\sqrt{\ln\left(\frac{2A_\text{MRR}}{\pi w_{zs}^2h_{ps,i}} \right)} }
		\Bigg/{\sqrt{2}\sigma_e^2}  \right) \text{d} h_{ps,i}.
\end{align}
Using \eqref{sb3}, $\mathbb{E}(h^2_{a1})=\mathbb{E}(h^2_{a2})=\mathbb{E}(h^2_{a})$ is obtained as:
\begin{align}
	\label{d7}
	&\mathbb{E}(h^2_{a1}) = \mathbb{E}(h^2_{1a1}h^2_{2a1}) = \mathbb{E}(h^2_{1a1})\mathbb{E}(h^2_{2a1}) \nonumber \\
	& = \int_0^\infty \frac{\alpha^\alpha h_{1a1}^{\alpha+1}}{\Gamma(\alpha)}e^{-\alpha h_{1a1}} \text{d} h_{1a1}
	\int_0^\infty \frac{\beta^\beta h_{2a1}^{\beta+1}}{\Gamma(\beta)}e^{-\beta h_{2a1}} \text{d} h_{2a1} \nonumber \\
	& = \frac{\Gamma(\alpha+2) \Gamma(\beta+2)}
	{\alpha^2 \beta^2 \Gamma(\alpha) \Gamma(\beta)}.
\end{align}
Finally, by substituting \eqref{d6}, \eqref{aq4} \eqref{sz1}, and \eqref{d7} in \eqref{aq3}, $\mathbb{E}(h^2_{si}|R_i)$ is obtained in \eqref{aq6}.

\section{} \label{AppB}
In the sensing phase, in most cases we have $R_i>>\sigma_e$. Under this condition, \eqref{sb5} is simplified as:
\begin{align} \label{z1}
	f_{r_i|R_i}(r_i) \simeq \frac{1}{\sqrt{2\pi \sigma_e^2}} \exp\left(- \frac{(r_i-R_i)^2}{2\sigma_e^2}  \right).
\end{align}
Using \eqref{sb5} and \eqref{z1}
\begin{align}
	\label{z2}
	\mathbb{E}(h_{ps,i}|R_i) &=\! \frac{2A_\text{MRR}}{\pi w_{zs}^2\sqrt{2\pi \sigma_e^2}}  \!
	\int_0^\infty    \!\!\!  
	\exp\left(\!-\!\frac{2r_i^2}{w_{zs}^2} \!- \frac{(r_i-R_i)^2}{2\sigma_e^2} \! \right)\! \text{d} r_i \nonumber \\
	& = A_1 e^{A_2 R_i^2}
\end{align}
where
\begin{align}
	\left\{
	\begin{array}{rl} 
		& A_1 = \frac{2A_\text{MRR}}{\pi w_{zs}^2 \sqrt{2A_3\sigma_e^2}}, ~~~
		A_2 = \frac{2 A_4^2 \sigma_e^2 - A_3}{2 A_3 \sigma_e^2}, \\
		& A_3 = \left( \frac{2}{w_{zs}^2} + \frac{1}{2\sigma_e^2}\right), ~~A_4=\frac{-1}{2\sigma_e^2}. \nonumber
	\end{array} \right.	
\end{align} 
Similarly, $\mathbb{E}(h^2_{ps,i}|R_i)$ is derived as:
\begin{align}
	\label{s6}
	& \mathbb{E}(h^2_{ps,i}|R_i) \simeq A_5  e^{A_6R_i^2 } 
	\left[ 1 - \text{erf}\left( \frac{A_4 R_i}{\sqrt{A_7}} \right) \right]
\end{align}
\begin{align}
	\left\{
	\begin{array}{rl} 
		& A_5 = \left( \frac{2A_\text{MRR}}{\pi w_{zs}^2}  \right)^2 
		\frac{ 1}{2 \sqrt{2A_7\sigma_e^2}},~~~
		A_6 = \frac{(2A_4 ^2\sigma_e^2-A_7)} {2A_7\sigma_e^2}, \\
		& A_7  = \left( \frac{4}{w_{zs}^2} + \frac{1}{2\sigma_e^2}\right). \nonumber
	\end{array} \right.	
\end{align}
Finally, substituting \eqref{z2} and \eqref{s6} in \eqref{ch7} and \eqref{aq6}, the PDF of $h_{si}$ conditioned on $R_i$ is derived in Proposition 1.

\section{}\label{AppC}
For large $w_z$, metric \eqref{ml1} is simplifies to:
\begin{align}
	\label{ml2}
	\hat{R}_i &=\underset{ R_i} {\text{arg min}}  
	~~~   (P_{rsi}-RP_t\mathbb{E}(h_{si}))^2 \nonumber \\
	& = \underset{ R_i} {\text{arg min}}  ~~~  RP_t\mathbb{E}^2(h_{si}) -2 P_{rsi} \mathbb{E}(h_{si}) .
\end{align}
Using the results of Proposition 1, metric \eqref{ml2} can be also simplified as:
\begin{align}
	\label{ml3}
	\hat{R}_i &=  \underset{ R_i} {\text{arg min}}  ~~~  
	RP_t\mathbb{A}_1  e^{2A_2 R_i^2}  -2 P_{rsi}    e^{A_2 R_i^2} .
\end{align}
By taking the derivative of \eqref{ml3} and setting it equal to zero, we get:
\begin{align}
	\label{ml4}
	\hat{R}_i =      \sqrt{\frac{1}{A_2}\ln\left( \frac{P_{rsi} }{RP_t\mathbb{A}_1}  \right)}.
\end{align}
Using \eqref{ml4}, the distribution of $\hat{R}_i$ is obtained as
\begin{align}
	\label{lm5}
	&f_{\hat{R}_i|R_i}(\hat{R}_i) \nonumber \\
	&= \frac{\text{d}}{\text{d} P_{rsi}}\text{Prob}\left[ 
	\sqrt{\frac{1}{A_2}\ln\left( \frac{P_{rsi} }{RP_t\mathbb{A}_1}  \right)}   < \hat{R}_i \bigg| R_i  \right] 
	\times \frac{\text{d} P_{rsi}}{\text{d} \hat{R}_i} \nonumber \\
	& = f_{P_{rsi}|R_i} \left( RP_t\mathbb{A}_1e^{A_2\hat{R}^2_i}  \right) \times \frac{\text{d} P_{rsi}}{\text{d} \hat{R}_i}.
\end{align}
Using \eqref{sd2}, \eqref{h1}, \eqref{h2} and \eqref{lm5}, after some manipulations, the PDF of $\hat{R}_i$ is derived in \eqref{lm6}.



\begin{thebibliography}{10} \balance
	\providecommand{\url}[1]{#1}
	\csname url@samestyle\endcsname
	\providecommand{\newblock}{\relax}
	\providecommand{\bibinfo}[2]{#2}
	\providecommand{\BIBentrySTDinterwordspacing}{\spaceskip=0pt\relax}
	\providecommand{\BIBentryALTinterwordstretchfactor}{4}
	\providecommand{\BIBentryALTinterwordspacing}{\spaceskip=\fontdimen2\font plus
		\BIBentryALTinterwordstretchfactor\fontdimen3\font minus
		\fontdimen4\font\relax}
	\providecommand{\BIBforeignlanguage}[2]{{%
			\expandafter\ifx\csname l@#1\endcsname\relax
			\typeout{** WARNING: IEEEtran.bst: No hyphenation pattern has been}%
			\typeout{** loaded for the language `#1'. Using the pattern for}%
			\typeout{** the default language instead.}%
			\else
			\language=\csname l@#1\endcsname
			\fi
			#2}}
	\providecommand{\BIBdecl}{\relax}
	\BIBdecl
	
	\bibitem{saeed2020cubesat}
	N.~Saeed, A.~Elzanaty, H.~Almorad, H.~Dahrouj, T.~Y. Al-Naffouri, and M.-S.
	Alouini, ``{CubeSat communications: Recent advances and future challenges},''
	\emph{IEEE Communications Surveys \& Tutorials}, vol.~22, no.~3, pp.
	1839--1862, 2020.
	
	\bibitem{2030_cube_market}
	\BIBentryALTinterwordspacing
	(2023) Cubesat market trends, demands, forecast to 2030. [Online]. Available:
	\url{https://straitsresearch.com/report/}
	\BIBentrySTDinterwordspacing
	
	\bibitem{NASA_nanoRack}
	\BIBentryALTinterwordspacing
	(2023) {TECHNICAL RESOURCES:} nanoracks and {NASA} document downloads.
	[Online]. Available: \url{https://nanoracks.com/resources/}
	\BIBentrySTDinterwordspacing
	
	\bibitem{ochoa2014deployable}
	D.~Ochoa, K.~Hummer, and M.~Ciffone, ``Deployable helical antenna for
	nano-satellites,'' 2014.
	
	\bibitem{4Mbps_Tx_marjet}
	\BIBentryALTinterwordspacing
	(2023) {CubeSatShop: ISIS TXS high data rate S-band transmitter}. [Online].
	Available:
	\url{https://www.cubesatshop.com/product/isis-txs-s-band-transmitter/}
	\BIBentrySTDinterwordspacing
	
	\bibitem{li2021advanced}
	L.~Li, Z.~Xuejiao, Z.~Jianhua, X.~Changzhi, and J.~Yi, ``Advanced space laser
	communication technology on cubesats,'' \emph{ZTE communications}, vol.~18,
	no.~4, pp. 45--54, 2021.
	
	\bibitem{kaushal2016optical}
	H.~Kaushal and G.~Kaddoum, ``{Optical communication in space: Challenges and
		mitigation techniques},'' \emph{IEEE communications surveys \& tutorials},
	vol.~19, no.~1, pp. 57--96, 2016.
	
	\bibitem{khalighi2014survey}
	M.~A. Khalighi and M.~Uysal, ``Survey on free space optical communication: A
	communication theory perspective,'' \emph{IEEE communications surveys \&
		tutorials}, vol.~16, no.~4, pp. 2231--2258, 2014.
	
	\bibitem{toyoshima2021recent}
	M.~Toyoshima, ``Recent trends in space laser communications for small
	satellites and constellations,'' \emph{Journal of Lightwave Technology},
	vol.~39, no.~3, pp. 693--699, 2021.
	
	\bibitem{carrasco2020space}
	A.~Carrasco-Casado and R.~Mata-Calvo, ``Space optical links for communication
	networks,'' in \emph{Springer Handbook of Optical Networks}.\hskip 1em plus
	0.5em minus 0.4em\relax Springer, 2020, pp. 1057--1103.
	
	\bibitem{chaudhry2022temporary}
	A.~U. Chaudhry and H.~Yanikomeroglu, ``Temporary laser inter-satellite links in
	free-space optical satellite networks,'' \emph{IEEE Open Journal of the
		Communications Society}, vol.~3, pp. 1413--1427, 2022.
	
	\bibitem{chaudhry2020free}
	A.~U. Chaudhry and H.~Yanikomerogl, ``Free space optics for next-generation
	satellite networks,'' \emph{IEEE Consumer Electronics Magazine}, vol.~10,
	no.~6, pp. 21--31, 2020.
	
	\bibitem{le2021throughput}
	H.~D. Le, P.~V. Trinh, T.~V. Pham, D.~R. Kolev, A.~Carrasco-Casado,
	T.~Kubo-Oka, M.~Toyoshima, and A.~T. Pham, ``{Throughput analysis for TCP
		over the FSO-based satellite-assisted internet of vehicles},'' \emph{IEEE
		Transactions on Vehicular Technology}, vol.~71, no.~2, pp. 1875--1890, 2021.
	
	\bibitem{le2023fso}
	H.~D. Le, H.~D. Nguyen, C.~T. Nguyen, and A.~T. Pham, ``{FSO-Based
		Space-Air-Ground Integrated Vehicular Networks: Cooperative HARQ With Rate
		Adaptation},'' \emph{IEEE Transactions on Aerospace and Electronic Systems},
	2023.
	
	\bibitem{walsh2022demonstration}
	S.~M. Walsh, S.~F. Karpathakis, A.~S. McCann, B.~P. Dix-Matthews, A.~M. Frost,
	D.~R. Gozzard, C.~T. Gravestock, and S.~W. Schediwy, ``{Demonstration of 100
		Gbps coherent free-space optical communications at LEO tracking rates},''
	\emph{Scientific Reports}, vol.~12, no.~1, p. 18345, 2022.
	
	\bibitem{carrasco2022development}
	A.~Carrasco-Casado, K.~Shiratama, D.~Kolev, P.~V. Trinh, F.~Ishola, T.~Fuse,
	and M.~Toyoshima, ``{Development and space-qualification of a miniaturized
		CubeSat’s 2-W EDFA for space laser communications},'' \emph{Electronics},
	vol.~11, no.~15, p. 2468, 2022.
	
	\bibitem{ishola2022characterization}
	F.~Ishola, A.~Carrasco-Casado, R.~Cordova, H.~Masui, Y.~Munemasa, P.~V. Trinh,
	T.~Fuse, H.~Tsuji, M.~Cho, and M.~Toyoshima, ``Characterization and
	comparison of {CubeSat} and drone platform jitter effects on laser beam
	pointing stability,'' 2022.
	
	\bibitem{kolev2023latest}
	D.~R. Kolev, A.~Carrasco-Casado, P.~V. Trinh, K.~Shiratama, F.~Ishola,
	H.~Kotake, J.~Nakazono, Y.~Saito, H.~Kunimori, T.~Kubooka \emph{et~al.},
	``Latest developments in the field of optical communications for small
	satellites and beyond,'' \emph{Journal of Lightwave Technology}, 2023.
	
	\bibitem{trinh2021experimental}
	P.~V. Trinh, A.~Carrasco-Casado, T.~Okura, H.~Tsuji, D.~R. Kolev, K.~Shiratama,
	Y.~Munemasa, and M.~Toyoshima, ``Experimental channel statistics of
	drone-to-ground retro-reflected fso links with fine-tracking systems,''
	\emph{IEEE Access}, vol.~9, pp. 137\,148--137\,164, 2021.
	
	\bibitem{born2018all}
	B.~Born, I.~R. Hristovski, S.~Geoffroy-Gagnon, and J.~F. Holzman, ``All-optical
	retro-modulation for free-space optical communication,'' \emph{Optics
		Express}, vol.~26, no.~4, pp. 5031--5042, 2018.
	
	\bibitem{quintana2021high}
	C.~Quintana, Q.~Wang, D.~Jakonis, O.~Oberg, G.~Erry, D.~Platt, Y.~Thueux,
	G.~Faulkner, H.~Chun, A.~Gomez \emph{et~al.}, ``A high speed retro-reflective
	free space optics links with {UAV},'' \emph{Journal of Lightwave Technology},
	vol.~39, no.~18, pp. 5699--5705, 2021.
	
	\bibitem{degnan2023tutorial}
	J.~J. Degnan, ``A tutorial on retroreflectors and arrays used in satellite and
	lunar laser ranging,'' in \emph{Photonics}, vol.~10, no.~11.\hskip 1em plus
	0.5em minus 0.4em\relax MDPI, 2023, p. 1215.
	
	\bibitem{moon2023performance}
	H.-J. Moon, C.-B. Chae, and M.-S. Alouini, ``Performance analysis of passive
	retro-reflector based tracking in free-space optical communications with
	pointing errors,'' \emph{IEEE Transactions on Vehicular Technology}, 2023.
	
	\bibitem{moon2023pointing}
	H.-J. Moon, C.-B. Chae, K.-K. Wong, and M.-S. Alouini,
	``Pointing-and-acquisition for optical wireless in {6G}: {From} algorithms to
	performance evaluation,'' \emph{IEEE Communications Magazine}, 2023.
	
	\bibitem{bashir2021optimal}
	M.~S. Bashir and M.-S. Alouini, ``Optimal power allocation between beam
	tracking and symbol detection channels in a free-space optical communications
	receiver,'' \emph{IEEE Transactions on Communications}, vol.~69, no.~11, pp.
	7631--7646, 2021.
	
	\bibitem{tsai2023angle}
	M.-C. Tsai, M.~S. Bashir, and M.-S. Alouini, ``Angle-of-arrival estimation of
	narrow gaussian beams for mobile fso platforms,'' \emph{arXiv preprint
		arXiv:2307.16002}, 2023.
	
	\bibitem{bashir2020adaptive}
	M.~S. Bashir and M.-S. Alouini, ``Adaptive acquisition schemes for
	photon-limited free-space optical communications,'' \emph{IEEE Transactions
		on Communications}, vol.~69, no.~1, pp. 416--428, 2020.
	
	\bibitem{10299800}
	M.~T. Dabiri and M.~Hasna, ``Performance analysis of modulating retroreflector
	array for {UAV-based FSO} links,'' \emph{IEEE Communications Letters}, pp.
	1--1, 2023.
	
	\bibitem{dabiri2022modulating}
	M.~T. Dabiri, M.~Rezaee, L.~Mohammadi, F.~Javaherian, V.~Yazdanian, M.~O.
	Hasna, and M.~Uysal, ``Modulating retroreflector based free space optical
	link for {UAV}-to-ground communications,'' \emph{IEEE Transactions on
		Wireless Communications}, vol.~21, no.~10, pp. 8631--8645, 2022.
	
	\bibitem{ghassemlooy2019optical}
	Z.~Ghassemlooy, W.~Popoola, and S.~Rajbhandari, \emph{{Optical wireless
			communications: system and channel modelling with
			Matlab{\textregistered}}}.\hskip 1em plus 0.5em minus 0.4em\relax CRC press,
	2019.
	
	\bibitem{andrews2005laser}
	L.~C. Andrews and R.~L. Phillips, ``Laser beam propagation through random
	media.''\hskip 1em plus 0.5em minus 0.4em\relax SPIE, 2005.
	
	\bibitem{saleh2019fundamentals}
	B.~E. Saleh and M.~C. Teich, \emph{Fundamentals of photonics}.\hskip 1em plus
	0.5em minus 0.4em\relax john Wiley \& sons, 2019.
	
	\bibitem{cannizzaro2019comparison}
	D.~Cannizzaro, M.~Zafiri, D.~Jahier~Pagliari, E.~Patti, E.~Macii, M.~Poncino,
	and A.~Acquaviva, ``{A comparison analysis of BLE-based algorithms for
		localization in industrial environments},'' \emph{Electronics}, vol.~9,
	no.~1, p.~44, 2019.
	
	\bibitem{Raician1}
	\BIBentryALTinterwordspacing
	(2016, May) Wolfram research (2010), ricedistribution, wolfram language
	function. [Online]. Available:
	\url{https://reference.wolfram.com/language/ref/RiceDistribution.html}
	\BIBentrySTDinterwordspacing
	
	\bibitem{karagiannidis2006bounds}
	G.~K. Karagiannidis, T.~A. Tsiftsis, and R.~K. Mallik, ``{Bounds for multihop
		relayed communications in Nakagami-m fading},'' \emph{IEEE Transactions on
		communications}, vol.~54, no.~1, pp. 18--22, 2006.
	
	\bibitem{Gam_func1}
	\BIBentryALTinterwordspacing
	Weisstein, eric w. "gamma function." from mathworld--a wolfram web resource.
	[Online]. Available: \url{https://mathworld.wolfram.com/GammaFunction.html}
	\BIBentrySTDinterwordspacing
	
\end{thebibliography}
\end{document}